\documentclass[UTF8,a4paper,twocolumn]{aastex631}

\usepackage{amsmath}

\usepackage{lineno}

\newcommand{\usigsfr}{M$_{\odot}$h$^{-1}$/Gyr (kpc/h)$^2$}

\submitjournal{ApJ}

\shorttitle{Disk formation in the TNG50 simulation}
\shortauthors{Wang \& Lilly}

\graphicspath{{fig/}{fig/example/}{fig/Inflow/}{fig/sfr_prof}}

\begin{document}

\title{The formation of star-forming disks in the TNG50 simulation}

\email{ecwang16@ustc.edu.cn}

\author[0000-0003-1588-9394]{Enci Wang}
\affil{Department of Physics, ETH Zurich, Wolfgang-Pauli-Strasse 27, CH-8093 Zurich, Switzerland}
\affil{CAS Key Laboratory for Research in Galaxies and Cosmology, Department of Astronomy, University of Science and Technology of China, Hefei 230026, People’s Republic of China}

\author[0000-0002-6423-3597]{Simon J. Lilly}
\affil{Department of Physics, ETH Zurich, Wolfgang-Pauli-Strasse 27, CH-8093 Zurich, Switzerland}

\begin{abstract}

We investigate the disk formation process in the TNG50 simulation, examining the profiles of SFR surface density ($\Sigma_{\rm SFR}$), gas inflow and outflow, and the evolution of the angular momentum of inflowing gas particles.  
The TNG50 galaxies tend to have larger star-forming disks, and also show larger deviations from exponential profiles in $\Sigma_{\rm SFR}$ when compared to real galaxies in the MaNGA (Mapping Nearby Galaxies at APO) survey.  
The stellar surface density of TNG50 galaxies show good exponential profiles, which is found to be the result of  strong radial migration of stars over time.  However, this strong radial migration of stars in the simulation produces flatter age profiles in TNG50 disks compared to observed galaxies. 
The star formation in the simulated galaxies is sustained by a net gas inflow and this gas inflow is the primary driver for the cosmic evolution of star formation, as expected from simple gas-regulator models of galaxies. 
There is no evidence for any significant loss of angular momentum for the gas particles after they are accreted on to the galaxy, which may account for the large disk sizes in the TNG50 simulation.  Adding viscous processes to the disks, such as the magnetic stresses from magneto-rotational instability proposed by \cite{Wang-22}, will likely reduce the sizes of the simulated disks and the tension with the sizes of real galaxies, and may produce more realistic exponential profiles. 

\end{abstract}


\section{Introduction}
\label{sec:1}

Massive star-forming galaxies in the local universe are widely seen to have a disky morphology \citep[e.g.][]{Simard-11, Meert-13}. The surface brightness of disk galaxies are typically composed of two components \citep[e.g.][]{Freeman-70, Kent-84, Allen-86, Weiner-01, Simard-11, Casasola-17}: a central spheroidal core and a highly flattened disk.   
The disk components are observed to have an nearly exponential broad-band photometric profile, extending over four scale lengths or more \citep[e.g.][]{Kent-85, Weiner-01, Pohlen-06, Meert-13}.  This characteristic exponential profile is found not only in the radial distribution of stars, as traced by optical-to-near infrared broad-band images, but also in the radial profile of the star formation rate (SFR) surface density \citep[$\Sigma_{\rm SFR}$; e.g.][]{Wyder-09, Gonzalez-Lopezlira-12, Gonzalez-Delgado-16, Casasola-17, Wang-19}, as traced by ultraviolet continuum, thermal infrared emission, and/or the H${\alpha}$ emission of ionized gas.

The most recent generation of cosmological hydrodynamical simulations, such as EAGLE \citep{Schaye-15} and IllustrisTNG \citep{Pillepich-18} represent a remarkable success in reproducing the development of the galaxy population over cosmic time \citep[see][and references therein]{Vogelsberger-20}. In general, such simulations successfully match many observational constraints, including  the galaxy colour bimodality, cold gas fractions, the statistical properties of galaxy morphology and other properties \citep{Trayford-15, Furlong-15, Nelson-18, Genel-18, Diemer-19, Donnari-19, Rodriguez-Gomez-19}.  For instance,  \cite{Nelson-18} found that, including the dust attenuation of stellar light, the simulated color distributions for massive galaxies ($M_*>10^9$M$_{\odot}$) in TNG100 are in excellent quantitative agreement with those of SDSS \citep[the Sloan Digital Sky Survey;][]{York-00} galaxies.  \cite{Rodriguez-Gomez-19} also showed that the optical morphologies of IllustrisTNG galaxies are in good agreement with the galaxies of Pan-STARRS \citep[Panoramic Survey Telescope and Rapid Response System;][]{Chambers-16} observations. 
However, it is not clear to what extent the simulations reproduce the detailed internal structure of disks that is seen in the observations, i.e. specifically the exponential form of the stellar disks and star-formation profiles.

The origin of the exponential disks in galaxies has been studied for more than half century, but is still not well understood.  Since the radial distribution of mass in a rotating disk is closely linked to the angular momentum distribution of the material in the disk, much attention has been paid to the initial angular momentum of the material that ultimately ends up in the disk, invoking conservation of specific angular momentum of each element during the collapse of proto-galaxy \citep[e.g.][]{Mestel-63, Freeman-70, Fall-80}. For instance, \cite{Freeman-70} showed that the angular momentum distribution of a self-gravitating exponential disk is remarkably similar to that of a uniformly rotating sphere of uniform density.    

This basic concept has been developed in many subsequent discussions, based on realistic N-body simulations of the formation of cosmic structure \citep[e.g.][]{Fall-80, Mo-98, Dutton-09}. These have ultimately tried to connect the specific angular momentum distribution of the baryonic material in galactic disks to the original distribution of that material, which is taken to match that of the dark matter particles in the host halos.  
On the other hand, by investigating the angular momentum transport of gas particles for individual disk galaxies in SPH simulations, \cite{Kaufmann-07} found that angular momentum of the particles can be lost in such simulations (presumably transported outwards), but that the severity of this
decreases substantially as the mass-resolution is increased and that this was therefore an artificial effect.  \cite{Kaufmann-07} claimed that, with $10^6$ gas and dark matter particles, disc particles lose only 10–20 per cent of their original angular momentum, and that exponential disks cannot be obtained. 

Another idea with a long history is that the formation of an exponential stellar disk is the result of secular evolution of the distribution of stars, after they are formed, as a result of bar formation \citep{Hohl-71, Foyle-08}, and/or by the scattering of stars by massive clumps \citep{Elmegreen-13, Wu-20}. Consistent with this, \cite{Herpich-17} suggested that the exponential stellar disk could be viewed as the maximum entropy state for the distribution of specific angular momentum of stars.  However, these ideas do not explain the exponential form of the SFR surface density in disk galaxies, since there is no clear reason why star-formation should follow a similar radial distribution to that of the long-lived stars.  

A third idea that has been explored over the years is that the exponential disk is produced by the operation of a viscous accretion disk \citep[e.g.][]{Lynden-Bell-74, Pringle-81, Lin-87, Yoshii-89, Wang-09, Wang-22, Wang-22b}. 
The viscosity in an accretion disk transports mass inwards and angular momentum outwards.  The viscosity in a gas disk can be produced by a number of processes, such as cloud-cloud collisions, turbulence of the gas disk from supernova feedback and/or the motions produced by gravitational instabilities of gas clouds \citep[e.g.][]{Lynden-Bell-74, Pringle-81, Ferguson-01, Stevens-16}, or as discussed below,  also by magnetic fields  \citep[][]{Wang-22b}.

The short gas depletion timescales of galaxies suggests that the gas in galaxies is being continually replenished from outside.   Assuming that the rate of star-formation of a galaxy depends on the available gas, and that outflows of gas are related to the star-formation rate, this leads to the idea that galaxies may be thought of as ``gas-regulator'' systems \citep{Lilly-13}, in which the gas-content of a galaxy continually adjusts itself to achieve a quasi-equilibrium between inflow, star-formation and outflow.  Variations of this idea have been explored in a number of papers \citep[e.g.][]{Bouche-10, Schaye-10, Dave-11, Lilly-13, Belfiore-19, Wang-21}.  

In parallel, multiple simulations based on different hydrodynamical codes show that the inflowing gas is almost co-planar and co-rotating with the gas disk regardless of its thermal history, at least at low redshifts \citep[e.g.][]{Kerevs-05, Stewart-11, Danovich-15, Stewart-17, Peroux-20, Trapp-22, Hafen-22, Gurvich-22}. Specifically, based on FIRE-2 \citep[Feedback In Realistic Environments;][]{Hopkins-18} simulation, \cite{Hafen-22} found that the inflowing gas becomes coherent and angular momentum-supported prior to accretion onto the disks.  On the other hand, the outflow of gas, driven by stellar winds and/or supernova (SN) explosions, is preferentially leave gas disks along the direction that is perpendicular to the disks \citep[e.g.][]{Nelson-19, Peroux-20, Trapp-22}. 

Motivated by these developments, we recently revisited the viscous disk idea in \cite{Wang-22}.   We focused, via a reverse-engineering approach, on what was required in order for a viscous co-planar gas disk to produce an exponential profile of star-formation in the disk.  In such a picture, the exponential stellar disk is then a natural outcome of the exponential form of the star-forming disk (although some subsequent migration or rearrangement of stars is not ruled out).  We showed that if galaxy disks can indeed be viewed as ``modified accretion disks'', in which the dominant gas flow is co-planar inflow within the disk but in which (unlike a classical accretion disk) gas is continually extracted from the disk due to star-formation (and any associated ex-planar outflows of gas), then the required viscous stresses $W(r)$ can be derived \textit{solely} 
from the steady-state star-formation rate profile $\Sigma_{\rm SFR}(r)$.    

We further argued that magneto-rotational instability (MRI) is an attractive and plausible source of the viscosity in galactic-scale disks. We showed that exponential $\Sigma_{\rm SFR}(r)$ profiles over several scale-lengths will be established if (and only if) there is a relation between the magnetic field strength and the $\Sigma_{\rm SFR}(r)$ that is of the precise type that has actually been indicated from observations of galaxies.    
It has been widely accepted for a long time that MRI is  likely to be an important source of the viscosity in classical accretion disks around compact objects, \citep[e.g.][]{Shakura-73, Blandford-NG, Balbus-91}.   \cite{Balbus-91} found that the combination of a negative gradient in the angular velocity with a weak magnetic field of any plausible astrophysical strength would lead to a dynamical instability, called the MRI.  Based on MHD simulations, \cite{Hawley-95} found that the transportation of angular momentum is dominated, for Keplerian disks, by the magnetic stress (or Maxwell stress), rather than by the kinetic stress (or Reynolds stress).  

As in all accretion disks, this viscosity causes an inward transport of mass within the galactic disk, which is required to sustain the exponential SFR surface density profile. There is also an outward transport of angular momentum within the gas disk associated with the loss of angular momentum as the gas moves down to the center of the disk.  The observed rotation curve means that, in this picture, the gas must lose a substantial fraction of its angular momentum as it flows inwards through the gas disk, e.g. losing 50\% of its specific angular momentum as it moves from a radius of 4 disk scalelengths down to 1 disk scalelength.   

In \cite{Wang-22b} we examined the expected metallicity gradients in this ``modified accretion disk" model of galaxy disks, and showed that this also is determined primarily by the $\Sigma_{\rm SFR}(r)$ profile, and is, perhaps counter-intuitively, \textit{independent} of the gas content or star-formation efficiency of the disk.   The model naturally produces a negative gradient of gas-phase metallicity due to  the progressive enrichment of gas by in-situ star formation in the disk as the gas flows down towards the center of the galaxy.  The expected profiles can quantitatively match observational results \citep[e.g.][]{Crockett-06, Magrini-07,Bresolin-12,Scarano-13}.  


In this paper, we focus on the formation of galaxy disks in hydro-simulations.  We take advantage of the publicly available  state-of-art simulation, TNG50 \citep[e.g.][]{Pillepich-18, Nelson-18}, which is the successor of the Illustris simulation \citep{Vogelsberger-14, Genel-14}, and appears to successfully reproduce many observational facts about the galaxy population\citep[e.g.][]{Diemer-19,Nelson-DR}.  Interestingly, \cite{Cannarozzo-22} focused on the early-type galaxies, and found that TNG100 generally produces an excellent agreement with the observations in the shape of stellar surface density, stellar age and metallicity. In this work, we instead focus on the disk galaxies and examine whether the simulation reproduces the exponential stellar disks, as well as the exponential star-forming disk.  

In addition to the general comparison between the output of simulations with observations, TNG50 and other similar simulations contain a wealth of information to address questions about how gas is accreted onto disks,  whether gas accretion or the pre-existing gas is dominant for sustaining star formation in galaxies, and how the angular momentum changes after the gas particles are accreted onto galaxies.  As the simulation of the highest resolution within the IllustrisTNG suite,  TNG50 is able to address these questions. 


The paper is organized as follows. In Section \ref{sec:2}, we will briefly introduce the TNG50 simulation, as well as the selection of the sample of simulated galaxies used in this work.  We examine whether the $\Sigma_{\rm SFR}$ profiles of simulated galaxies are close to exponential or not in Section \ref{sec:3}, and provide a quantitative comparison with the observational results from MaNGA galaxies.  Another focus of this work is to investigate the inflow and outflow of gas particles on the disks of individual galaxies, in order to examine whether the gas disk of different radii can be treated as a gas-regulator system or not in simulations. These results are shown in Section \ref{sec:4}.  In Section \ref{sec:5}, we will track individual inflowing gas particles to examine the change of their angular momentum after being accreted onto the gas disks.  We then summarize this work in Section \ref{sec:6}. 

Throughout the paper, we adopt a flat cold dark matter cosmology model with cosmological parameters derived from \cite{Planck-Collaboration-and-Ade-16}, i.e. $\Omega_{\Lambda,0}=$0.6911, $\Omega_{m,0}=$0.3089, $\Omega_{b,0}=$0.0486, and H$_0$=67.74 km s$^{-1}$Mpc$^{-1}$. This is the same as the one adopted in IllustrisTNG simulation \citep{Nelson-DR}. 

\section{The TNG50 simulation}
\label{sec:2}

\subsection{Brief introduction of TNG50} \label{sec:2.1}


As the successor to the original Illustris simulation\citep[e.g.][]{Vogelsberger-14, Sijacki-15}, IllustrisTNG \citep[e.g.][]{Springel-18,Nelson-18} is a suite of state-of-the-art magneto-hydrodynamic cosmological simulations with an updated physical model, run with the moving-mesh code {\tt AREPO} \citep{Springel-10}.  The implemented physical processes include star formation, stellar evolution, chemical enrichment, outflows driven by stellar feedback and etc \citep[see][for details]{Pillepich-18}. 

In IllustrisTNG, star formation and pressurization of the multi-phase ISM are implemented following the model of \cite{Springel-03}.   Specifically, stars form stochastically following the empirically defined Kennicutt–Schmidt relation, in gas clouds above a density threshold of $n_{\rm H}=0.1$ cm$^{-3}$.  The recipe of star formation driven kinetic winds is refined in several ways with respect to the approach in the Illustris simulation \citep{Vogelsberger-13, Torrey-14}. For instance, winds are injected isotropically in IllustrisTNG, and the initial speed of the wind particles is set to be redshift-dependent. The total energy release rate that is available to drive galactic winds, is set by the instantaneous star formation rate and the energy released by Type II SN per unit stellar mass that is formed. Both depend on spatial location within a galaxy and on time.  

With these improved treatments of physical processes, the output of IllustrisTNG has been shown to be consistent with a wide range of observational data in addition to those that were used to calibrate or tune the model \citep{Nelson-DR}. These test data include the galaxy stellar mass functions up to $z\sim4$ \citep{Pillepich-18}, the spatial clustering of red and blue galaxies up to tens of Mpc \citep{Springel-18}, the color-magnitude diagram of blue and red galaxies \citep{Nelson-18}, and the stellar sizes of star-forming and quiescent galaxies up to $z\sim2$ \citep{Genel-18}, the optical morphologies of galaxies \citep{Rodriguez-Gomez-19}.  These, and the publicly available output at the particle level, 
make the IllustrisTNG one of the best choices to study the formation of disks in galaxies within cosmological simulations.  

The IllustrisTNG project includes three simulation volumes with different resolutions. In this work, we use the highest-resolution version of IllustrisTNG50-1 (hereafter, TNG50 for short).  TNG50 simulation starts at $z=127$ and evolves down to $z=0$, with a box side length of  $\sim$50 Mpc. It initially contains $2160^{3}$ dark matter particles and $2160^{3}$ gas cells.  The mass of each dark matter particle is $3.07\times 10^{5}{\rm M}_{\odot}/h$, and the average gas cell mass is $5.74\times 10^{4}{\rm M}_{\odot}/h$. The softening length employed in TNG50 is 0.29 kpc for both dark matter and stellar particles, and an adaptive gravitational softening length for gas cells is adopted, with a minimum value of 0.074 kpc. In this work, the gas cells are also named as gas particles. 

In the released data, the halos and subhalos are identified by the {\tt FoF} and {\tt Subfind} algorithms \citep{Springel-05, Dolag-09}, and the merger tress are constructed by the {\tt SubLink} algorithms \citep{Rodriguez-Gomez-15}.  Simulated galaxies are associated with the subhalos, while the merger trees are useful to trace their histories of formation and evolution. 

\subsection{The sample selection from TNG50} \label{sec:2.2}

\begin{figure*}[htb]
  \centering
\includegraphics[width=0.45\textwidth]{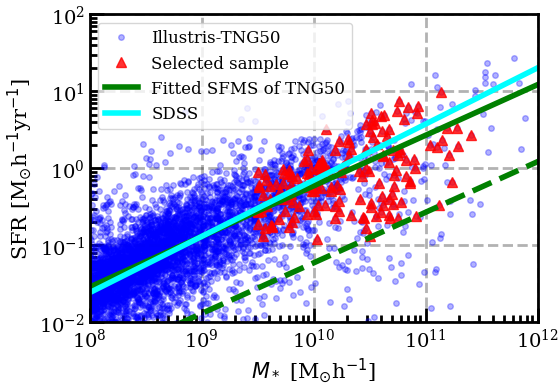}
\includegraphics[width=0.45\textwidth]{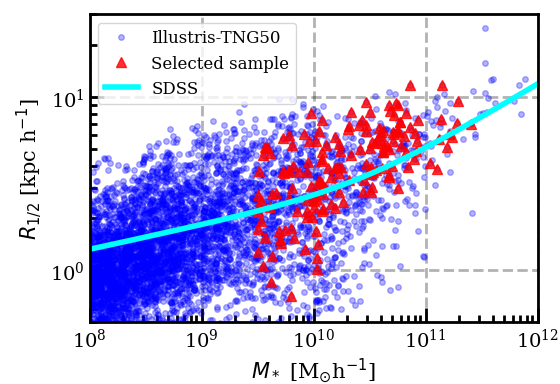}
  \caption{ Left panel: The stellar mass-SFR relation for the galaxies in the TNG50 simulation at redshift of $z=0$.  We perform a linear fitting to the star-formation main sequence for the TNG50 galaxies iteratively, which is shown in green solid line. The green dashed line is the defined demarcation line of star-forming galaxies and quenched galaxies, and is 1 dex below the green solid line.  For comparison, we show the SFMS of SDSS galaxies in cyan solid line \citep{Bluck-16}.  The selected subsample, investigated in the present work, is shown in red triangles. 
  Right panel: The stellar mass-size relation for the TNG50 galaxies.  The selected subsample  is shown in red triangles.  For comparison, we show the mass-size relation of SDSS late-type galaxies in cyan line \citep{Shen-03}, multiplied by 1.4 to account for the use of circularized radii \citep{Furlong-17}.  }
  \label{fig:1}
\end{figure*}

Since star-forming galaxies almost always show well-defined disks, we select the sample of galaxies to be studied from the SFR-$M_*$ diagram. The left panel of Figure \ref{fig:1} shows the SFR-$M_*$ relation for TNG50 galaxies at redshift of $z=0$.   Throughout the paper, the stellar mass of the simulated galaxies is defined as the total mass of the member stellar particles within the radius at which the mean surface brightness profile (computed from all member stellar particles) drops below the limit of 20.7 mag arcsec$^{-2}$ in the K band \citep[the {\tt SubhaloStellarPotometricMassinRad} of galaxy catalog from TNG50 data release;][]{Nelson-DR}.  Here, the SFR is defined as the total SFR of the cells within twice the stellar half-mass radius (the {\tt SubhaloSFRinRad} of galaxy catalog from TNG50 data release). These definitions of SFR and stellar mass are only used in the current work to define the sample selection.  We note that different definitions of SFR are adopted later when calculating the $\Sigma_{\rm SFR}$ of galaxies, as specified in Section \ref{sec:3} and Section \ref{sec:4}. 

As can be seen, the TNG50 galaxies are found on a tight sequence in the SFR-$M_*$ diagram, reproducing the so-called star formation main sequence \citep[SFMS;][]{Brinchmann-04, Daddi-07, Elbaz-11} of real galaxies.  We perform a linear fitting to the SFMS of the TNG50 galaxies iteratively, and this is shown in green solid line. Specifically, we first fit a straight line of SFR-$M_*$ relation for all the TNG50 galaxies. Then we re-fit the relation by excluding the galaxies that are 1 dex below the fitted SFMS. Repeating the above process for a few times, we then obtain a fitted SFMS, shown in green solid line on the left panel of Figure \ref{fig:1}.  The green dashed line is the demarcation line of star-forming and quenched galaxies, which is 1 dex below the finally fitted SFMS.  For comparison, we show the observed SFMS of real SDSS galaxies as the cyan solid line, taken from \cite{Bluck-16}. The fitted SFMS of TNG50 galaxies appears to be in excellent agreement with that of observations. 

In this work, in order to select a representative sample of manageable size for detailed analysis, we select four subsamples of star-forming galaxies based on the SFR-$M_*$ diagram.   Specifically, we randomly select 50 star-forming TNG50 galaxies, excluding galaxies with significant mergers (mass-ratio greater than 0.1) since redshift of 0.4, in each three stellar mass intervals: $9.5<\log M_*/({\rm M}_{\odot}h^{-1})<10.0$, $10.0<\log M_*/{\rm M}_{\odot}h^{-1})<10.5$ and 
$10.5<\log M_*/({\rm M}_{\odot}h^{-1})<11.0$. 
In a fourth mass bin, $11.0<\log M_*/({\rm M}_{\odot}h^{-1})<11.5$, where there are fewer galaxies, we take all the star-forming galaxies, but exclude the mergers as before.  This results in 11\footnote{One galaxy that matches our selection criteria is further excluded as it frequently caused problems in computing the net inflow rate on TNG {\tt JupiterLab}. This may be due to the fact that this galaxy contains too many gas particles. We therefore do not include this one galaxy here and also in the following analysis. } galaxies.  In the remaining analysis of this work, we will focus on only these 161 selected sample galaxies, which are shown with red triangles in Figure \ref{fig:1}.  

The right panel of Figure \ref{fig:1} shows the mass-size relation of TNG50 galaxies at redshift of $z=0$. The size used here is the half-stellar mass radius, denoted as $R_{1/2}$.  For comparison, we show the mass-size relation of late-type galaxies of SDSS as a cyan solid line, taken from \cite{Shen-03}. However, the circularized radii\footnote{The circularized radius is defined as $\sqrt{b/a}$ times the major-axis size, where a and b the minor- and major-axes.} used by \cite{Shen-03} are typically a factor of 1.4 smaller than uncircularized radii, or the major-axis radii \citep{van-der-Wel-14, Furlong-15}.  We therefore shift the mass-size relation of \cite{Shen-03} vertically by a factor of 1.4 in the right panel of Figure \ref{fig:1}, to account for this effect. 

We find overall trend of the mass-size relation of TNG50 is in quite good agreement with that from observations, over four orders of magnitude in stellar mass, but that there is an offset of about 0.11 dex at high mass end. Similarly, \cite{Genel-18} investigated the evolution of the mass-size relation for galaxies of TNG100, and found that a quantitative comparison of the projected $r$-band sizes in TNG100 and in observations \citep{van-der-Wel-14}, also shows overall agreement to within 0.25 dex from $0.0<z<0.2$ but with the TNG100 galaxies having systematically larger sizes than the real galaxies with $M_*> 10^{10.0}{\rm M}_{\odot}h^{-1}$, which is similar to what we found for TNG50 galaxies in the right panel of Figure \ref{fig:1}.

\section{Do the simulations form exponential disks? }
\label{sec:3}

\subsection{The profiles of SFR surface density} \label{sec:3.1}

\begin{figure*}
  \centering
\includegraphics[width=0.42\textwidth]{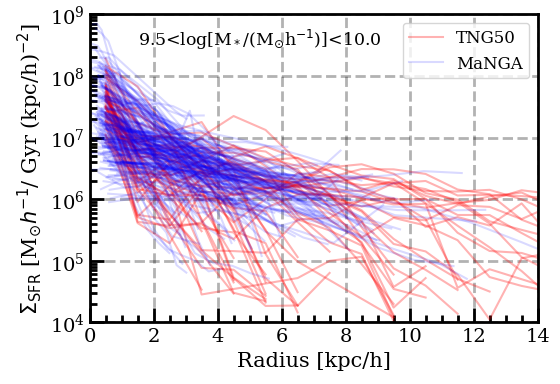}
\includegraphics[width=0.42\textwidth]{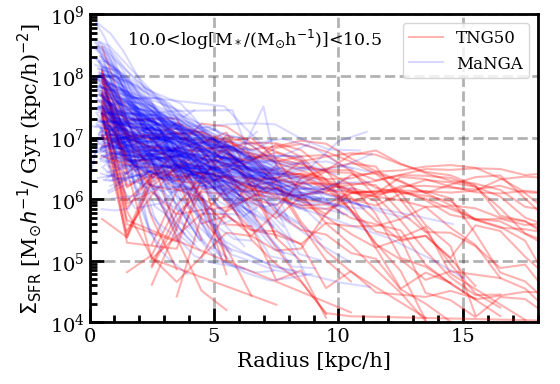}
\includegraphics[width=0.42\textwidth]{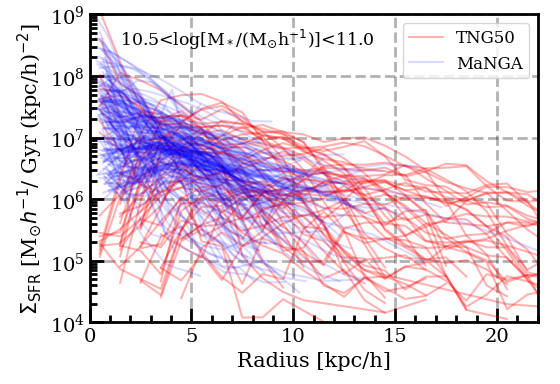}
\includegraphics[width=0.42\textwidth]{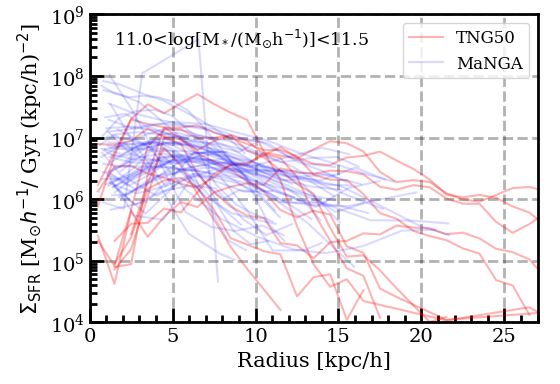}
  \caption{ The mean SFR surface density profiles of TNG50 galaxies (red lines) and MaNGA galaxies (blue lines). The galaxies are separated into 4 stellar mass bins, as denoted in each panels.  }
  \label{fig:2}
\end{figure*}

\begin{figure*}
  \centering
\includegraphics[width=0.42\textwidth]{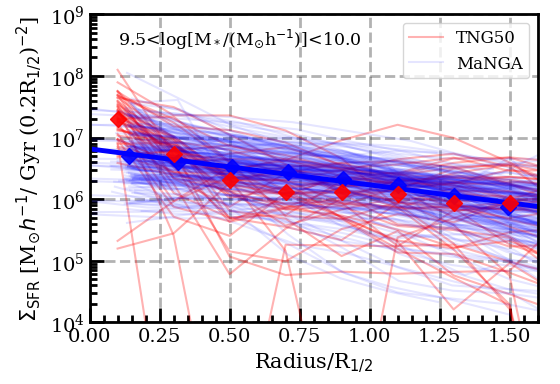}
\includegraphics[width=0.42\textwidth]{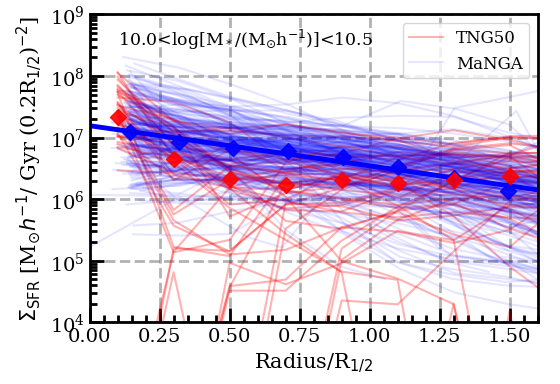}
\includegraphics[width=0.42\textwidth]{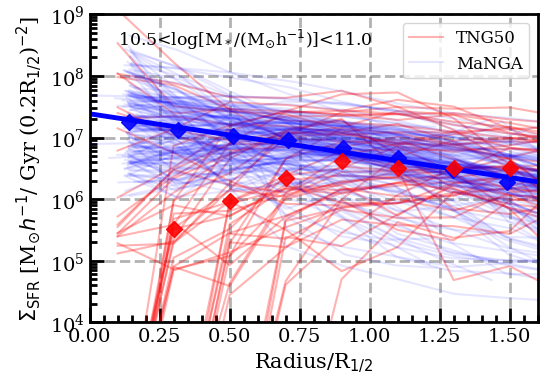}
\includegraphics[width=0.42\textwidth]{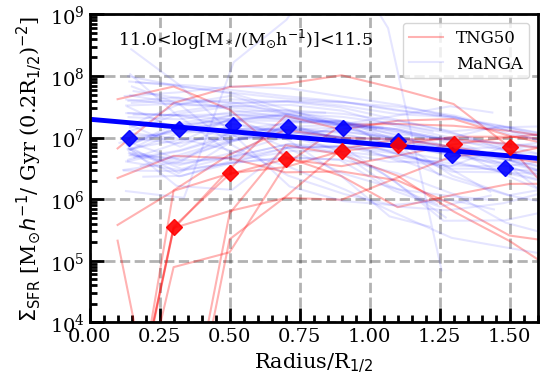}
\caption{ The same as Figure \ref{fig:2}, but showing the  profiles as a function of the normalized radius r/R$_{1/2}$ (or the effective radius for MaNGA galaxies). We also plot a normalized SFR surface density, which is computed as the SFR in an area that itself scales as R$_{1/2}^2$. This ensures that the visual integration of a profile on the $\Sigma_{\rm SFR}$-radius diagram reflects the actual integrated quantity in physical terms \citep[see also][]{Wang-19}. In each panel, the blue and red diamonds shows the median $\Sigma_{\rm SFR}$ of that stellar mass bin for the MaNGA and TNG50 galaxies, respectively.  The median $\Sigma_{\rm SFR}$ profiles of MaNGA galaxies are close to exponential functions, and we therefore fit an exponential function to the MaNGA median $\Sigma_{\rm SFR}$ profile shown in the blue solid line of each panel. 
}
  \label{fig:3}
\end{figure*}


As mentioned in the introduction, the $\Sigma_{\rm SFR}$ profiles of star-forming galaxies typically have an exponential form, which may well naturally account for the exponential form of the stellar disks. Specifically, \cite{Wang-22} looked at the deviations of individual $\Sigma_{\rm SFR}(r)$ profiles of MaNGA \citep[Mapping Nearby Galaxies at APO;][]{Bundy-15} star-forming galaxies when compared with pure exponential profiles.  In the current analysis, the $\Sigma_{\rm SFR}(r)$ of MaNGA galaxies at a given radius was calculated as the mean $\Sigma_{\rm SFR}$ of pixels within the corresponding annulus, rather then the median, in order to match our treatment of the simulated TNG50 galaxies. 

It was found that more than half (or 86\%) of galaxies with the rms deviations or less than 0.1 dex (or 0.2 dex).  Observationally, exponential star-forming disks seem to be a very common feature of galaxies, especially for galaxies with low-to-intermediate stellar mass \citep[$\log M_*/{\rm M}_{\odot}<10.7$;][]{Wang-22}.  Therefore, in this subsection, we investigate the $\Sigma_{\rm SFR}$ profiles of TNG50 galaxies, and examine whether the exponential form is found in them, as compared to real galaxies.   


The red lines in Figure \ref{fig:2} show the mean $\Sigma_{\rm SFR}$ profiles of individual galaxies for the selected TNG50 sample based on the $z=0$ snapshot.   In calculating the radial profiles of $\Sigma_{\rm SFR}(r)$ for each simulated galaxy, we first determine the center of the galaxy, and then determine the 3-d orientation of the disk by minimizing the mean absolute perpendicular distance of all star particles from the plane of the disk, testing all possible orientations of the disk in space that are centered on the galactic center.  
We stress that, throughout this paper, the orientation of the disk is always determined based on  the distribution of stellar particles.  

We can then obtain the face-on distribution of stellar particles for each individual galaxy. By radially binning the stellar particles into a set of annuli, the mean $\Sigma_{\rm SFR}(r)$ of each TNG50 galaxies in Figure \ref{fig:2} is then computed from the {\it initial mass} of stellar particles that are formed within each annulus within the last 100 Myr and is therefore an average  $\Sigma_{\rm SFR}$ on a 100 Myr timescale.  This reduces the shot noise.  
Based on this method, we have estimated that the signal-noise ratio of $\Sigma_{\rm SFR}$ induced by shot noise is typically greater than 3.0 for $\Sigma_{\rm SFR}\sim 10^{5}$ \usigsfr\ with a radial bin of 1 kpc at 8 kpc, given the typical mass of $5.7\times 10^4$M$_{\odot}h^{-1}$ for gas particle.  

For comparison, we show the mean $\Sigma_{\rm SFR}$ for a well-defined sample of MaNGA star-forming galaxies as the blue lines in Figure \ref{fig:2}.  These are taken from \cite{Wang-19}. Here we only briefly describe the sample definition and the calculation of $\Sigma_{\rm SFR}$, and refer the reader to \cite{Wang-19} for further details. This galaxy sample is originally selected from the SDSS Data Release 14 \citep{Abolfathi-18}, excluding quenched galaxies, mergers, irregulars, and heavily disturbed galaxies.  This results in a sample of 976 MaNGA star-forming galaxies, which is a good representation of normal SFMS galaxies.  The SFR for MaNGA galaxies is calculated based on the dust-corrected H$\alpha$ luminosity with the \cite{Kennicutt-98} star formation law and \cite{Chabrier-03} initial mass function (IMF).   The spatial coverage of MaNGA galaxies is typically larger than 1.5 effective radius. The $\Sigma_{\rm SFR}$ profiles of MaNGA galaxies are corrected for the projection effect, based on the minor-to-major axis ratio from the NSA catalog \citep{Blanton-11}.  The spatial resolution of MaNGA observation is 1-2 kpc, so in generating the profiles for the TNG50 galaxies, a radial bin width of 1 kpc was chosen to minimize the effect of this spatial resolution. We have tested the effect of bin size, and find that our main result is unchanged when using a bin size of 0.5 kpc and 2 kpc. 

As shown in Figure \ref{fig:2}, although the individual $\Sigma_{\rm SFR}$ profiles appear to be noisy, we find that the $\Sigma_{\rm SFR}$ profiles of TNG50 galaxies are overall overlapped with those of MaNGA 
galaxies for all the four stellar mass bins, suggesting a good overall consistency.  However, when looking in more detail, the shapes of the $\Sigma_{\rm SFR}$ profiles of the simulated TNG50 galaxies generally appear to be flatter and more extended than those of real MaNGA galaxies.   This is in line with the fact that TNG50 galaxies are overall somewhat larger than the real galaxies in the mass bins we considered, as was shown in the right panel of Figure \ref{fig:1}. 

To reduce the effect of different sizes between different galaxies and different samples, we replot the $\Sigma_{\rm SFR}$ profiles of TNG50 galaxies and MaNGA galaxies in Figure \ref{fig:3}, by normalizing the radius with the half-stellar-mass radius (or the effective radius for MaNGA galaxies).  We also re-normalized the $\Sigma_{\rm SFR}$, so that it is computed as the SFR within an area that itself scales as R$_{1/2}^2$. This ensures that the visual integration of a profile on the $\Sigma_{\rm SFR}$-radius diagram reflects the actual integrated quantity in physical terms \citep[see also][]{Wang-19}.  As shown in Figure \ref{fig:3}, 
the profiles of all the galaxies tend to be more parallel after the normalization and make the exponential form easier to see on the diagram.
In addition, for each stellar mass bins, we also show the median $\Sigma_{\rm SFR}$ profile of the populations with red and blue diamonds for the TNG50 and MaNGA samples, respectively.  These median profiles can be treated as a representative profile that captures the general features of the individual profiles within the corresponding stellar mass bin.  

As can be seen in Figure \ref{fig:3}, the median $\Sigma_{\rm SFR}$ profiles of MaNGA galaxies can be well-fitted by an exponential function for all the stellar mass bins except for the very highest ones \citep[also see][]{Wang-19}.  However, the median $\Sigma_{\rm SFR}$ profiles of TNG50 galaxies show quite large deviations from an single exponential function, and this is true for all the stellar mass bins examined.  We see a gradual change of the median $\Sigma_{\rm SFR}$ profile of TNG50 galaxies with increasing stellar mass: the median $\Sigma_{\rm SFR}$ profiles of two lowest stellar mass bin are clearly centrally-peaked ($<0.4R_{1/2}$), and become nearly flat out to the radius of 1.6$R_{1/2}$ or more; the median $\Sigma_{\rm SFR}$ profiles of the two highest stellar mass bins clearly show, in contrast, centrally-suppressed star formation ($<0.6R_{1/2}$), and become flat up to the radius of 1.6$R_{1/2}$ as we considered here.

It is clear that the $\Sigma_{\rm SFR}$ profiles of TNG50 galaxies show larger deviations from the pure exponential function with respect to those of MaNGA galaxies.  We will further explore this in the next section. 

\subsection{Quantifying the deviation from the exponential profiles } \label{sec:3.2}

\begin{figure*}[htb]
  \centering
\includegraphics[width=0.42\textwidth]{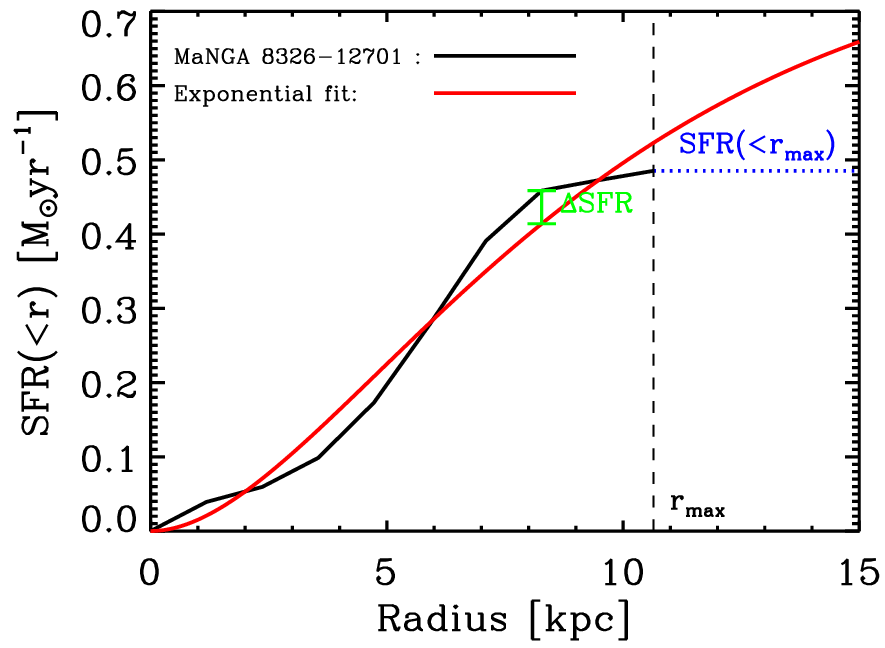}
\includegraphics[width=0.42\textwidth]{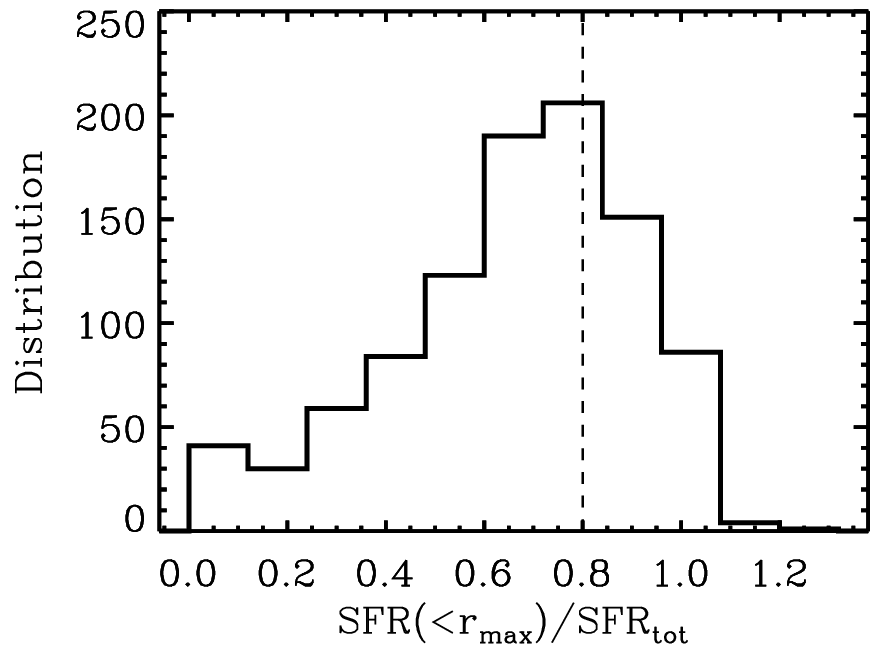}
  \caption{ Left panel:  An example to illustrate the definitions of $r_{\rm max}$, SFR($<r_{\rm max}$), $\Delta$SFR and SFR$_{\rm tot}$.  The black solid line shows the cumulative SFR as a function of radius for an individual MaNGA galaxy (8326-12701), and the red solid line shows the fit of the cumulative SFR with an exponential $\Sigma_{\rm SFR}$ profile.   Therefore, the $\Delta$SFR is defined as the maximum (absolute) deviation of the black line to the red fitted line, shown in green line segment.  The r$_{\rm max}$ shows the maximum radius of the coverage in MaNGA observation, and the SFR($<r_{\rm max}$) is the SFR within the radius of r$_{\rm max}$.  The SFR$_{\rm tot}$ is the fitted total SFR, i.e. SFR($<+\infty$).
  Right panel: The distribution of the ratio of SFR($<r_{\rm max}$) to SFR$_{\rm tot}$ for MaNGA galaxies. This distribution is peaked at  SFR($<r_{\rm max}$)/SFR$_{\rm tot}\sim$ 0.8. We therefore perform a similar analysis for TNG50 galaxies, defining the $r_{\rm max}$ as the radius within which the 80\% of star formation is enclosed.   
}
  \label{fig:4}
\end{figure*}

\begin{figure*}[htb]
  \centering
\includegraphics[width=0.42\textwidth]{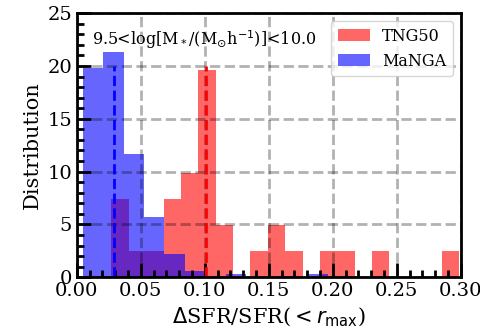}
\includegraphics[width=0.42\textwidth]{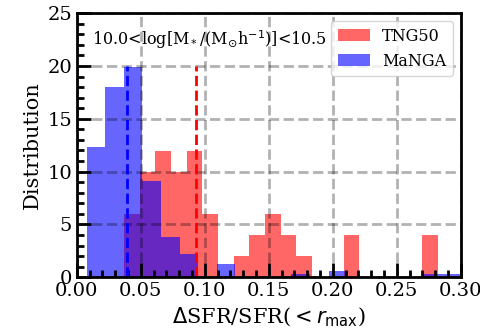}
\includegraphics[width=0.42\textwidth]{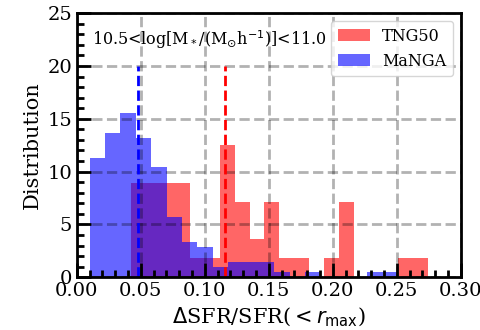}
\includegraphics[width=0.42\textwidth]{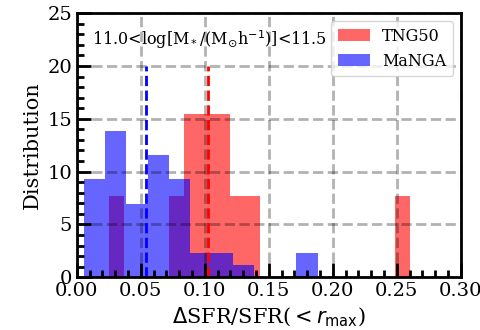}
  \caption{The distribution of $\Delta$SFR/SFR($<r_{\rm max}$) for TNG50 (red histograms) and MaNGA (blue histograms) galaxies. In each panel, the red and blue vertical dashed lines indicate the median values of $\Delta$SFR/SFR($<r_{\rm max}$) for TNG50 and MaNGA, respectively. As above, galaxies are separated into the four stellar mass bins.  }
  \label{fig:5}
\end{figure*}

\begin{figure*}[htb]
  \centering
\includegraphics[width=0.42\textwidth]{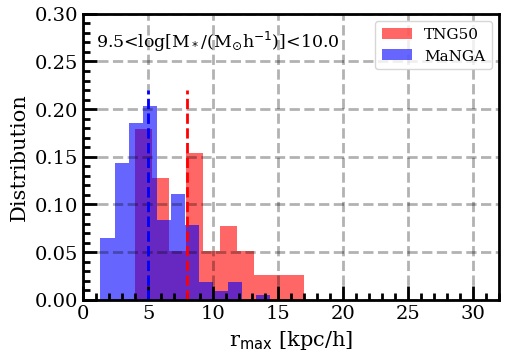}
\includegraphics[width=0.42\textwidth]{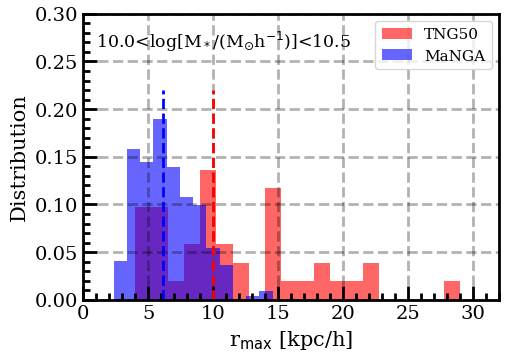}
\includegraphics[width=0.42\textwidth]{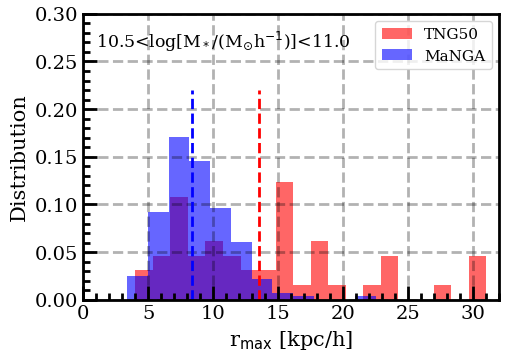}
\includegraphics[width=0.42\textwidth]{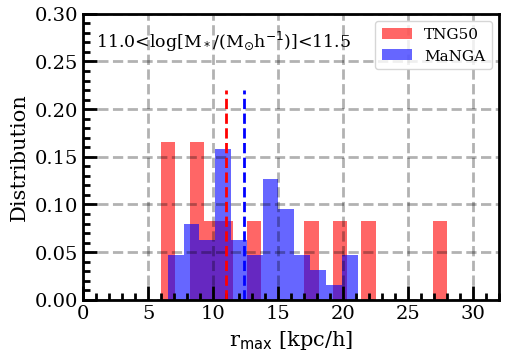}
  \caption{The distribution of $r_{\rm max}$ for TNG50 (red histograms) and MaNGA (blue histograms) galaxies.  The $r_{\rm max}$ roughly corresponds to the radius within which 80\% of star formation is enclosed for each galaxy. The red and blue vertical dashed lines indicate the median values of $r_{\rm max}$ for TNG50  and MaNGA, respectively.    As above, galaxies are separated into the four stellar mass bins.  }
  \label{fig:6}
\end{figure*}

In this subsection, we try to further quantify the deviations of the $\Sigma_{\rm SFR}(r)$ profiles of TNG50 and MaNGA galaxies from the pure exponential function.  To do this, we define a set of quantities for individual galaxies based on their $\Sigma_{\rm SFR}(r)$ profile in the following way.  

For an individual MaNGA galaxy, we first calculate the cumulative profile of SFR($<r$), i.e. the total SFR that is enclosed within the (de-projected) radius $r$.  
This cumulative SFR profile therefore stops at the radius corresponding to the coverage of MaNGA observation for that individual galaxy, which is denoted as $r_{\rm max}$.  
Then we fit the observed cumulative SFR profile with the cumulative function of an pure exponential $\Sigma_{\rm SFR}(r) \propto \exp(-r/h_{\rm R})$, which can be written as \citep[also see equation 8 in][]{Wang-22}: 
\begin{equation} \label{eq:1}
    {\rm SFR}(<r) =  {\rm SFR_{\rm tot}} \cdot [1-(r/h_{\rm R}+1)\cdot \exp(-r/h_{\rm R})], 
\end{equation}
where SFR$_{\rm tot}$ is the {\it total} SFR of the whole galaxy, and $h_{\rm R}$ is the scalelength of the idealised exponential $\Sigma_{\rm SFR}(r)$. For illustration, we show an example (MaNGA ID: 8326-12701) in the left panel of Figure \ref{fig:4}, where the black line is the observed SFR($<r$) and the red line is the best-fit curve of Equation \ref{eq:1}.  

We then define $\Delta$SFR as the maximum absolute deviation of the observed SFR($<r$) curve to the fitted one, which is shown as the green line segment in the left panel of Figure \ref{fig:4}.  The ratio of $\Delta$SFR to SFR($<r_{\rm max}$) is then a good parameter to quantify the deviation of $\Sigma_{\rm SFR}$ from an pure exponential function.  It is a measure of the fraction of the star-formation in the galaxy (or at least of that which is spatially observable) does \textit{not} follow a perfect exponential profile.

As just noted, the finite area of the MaNGA data do not cover all of the star-formation in these galaxies.  The right panel of Figure \ref{fig:4} shows the distribution of the ratio of SFR($<r_{\rm max}$) to the fitted SFR$_{\rm tot}$.  The SFR($<r_{\rm max}$)/SFR$_{\rm tot}$ peaks at $\sim$0.8, which means that the MaNGA observations have typically covered $\sim$80\% of the total star formation in these galaxies.   In order to apply as similar a procedure to TNG50 galaxies as possible, we therefore define an artificial $r_{\rm max}$ for the TNG50 galaxies to be the radius within which 80\% of their total star formation is enclosed.    We then measure the $\Delta$SFR/SFR($<r_{\rm max}$) for TNG50 galaxies as for the MaNGA ones.

Figure \ref{fig:5} shows the distributions of $\Delta$SFR/SFR($<r_{\rm max}$) for TNG50 galaxies (red histograms) and MaNGA galaxies (blue histograms) for the four stellar galactic mass bins.  As shown, the $\Delta$SFR/SFR($<r_{\rm max}$) of MaNGA galaxies is systematically less than that of the TNG50 galaxies across the full range of stellar mass that we considered.  Specifically, the median $\Delta$SFR/SFR($<r_{\rm max}$) of MaNGA for the four stellar mass bins are 0.029, 0.039, 0.048 and 0.054 with increasing mass, while the median $\Delta$SFR/SFR($<r_{\rm max}$) of TNG50 galaxies are 0.101, 0.093, 0.116 and 0.103, respectively.  In general, the median $\Delta$SFR/SFR($<r_{\rm max}$) of TNG50 galaxies are typically a factor of 2-3 larger as those of MaNGA galaxies, indicating that TNG50 galaxies show significantly larger deviation of their $\Sigma_{\rm SFR}$ profiles from the pure exponential function than do MaNGA galaxies.  

In a similar way, we show the distribution of $r_{\rm max}$ for the TNG50 galaxies and MaNGA galaxies in the four stellar mass bins in Figure \ref{fig:6}.  Since $r_{\rm max}$ (roughly) corresponds to the radius that contains 80\% of star formation, the value of $r_{\rm max}$ can be taken to reflect a measure of the size of the star-forming disks.  As shown in Figure \ref{fig:6}, TNG50 galaxies typically have larger star-forming disk than the MaNGA galaxies for all the stellar mass bins except the highest one.  Specifically, the median $r_{\rm max}$ of MaNGA galaxies are 5.0, 6.1, 8.4, and 12.3 kpc$/h$ with increasing stellar mass for the four mass bins, while the median  $r_{\rm max}$ of TNG50 galaxies are 8.0, 10.0, 13.5 and 11.0 kpc$/h$ with increasing mass, respectively.   Except in the highest mass bin, TNG50 galaxies typically have larger star-forming disks (as defined here by our $r_{\rm max}$ parameter) by a factor of $\sim$1.6 (or 0.2 dex larger) than MaNGA galaxies with $\log M_*/({\rm M_{\odot}}h^{-1})<11.0$.  
In fact, it is quite common that cosmological simulations produce larger galaxies than the observations, not just in IllustrisTNG  \citep[also see][]{Genel-18}.  For instance, by investigating the sizes of galaxies in the EAGLE simulation \citep{Schaye-15}, \cite{Furlong-15} found that the predicted mass-size relation is systematically shifted with respect to that of observations, with the simulated galaxies 0.2 dex larger than real ones at a given stellar mass.  By comparing the size of galaxies from Illustris and SDSS, \cite{Bottrell-17} found the simulated galaxies are also roughly twice as large (0.3 dex) on average as real galaxies when matching stellar masses. 

The general conclusion from this section is that star-formation is occurring too far out in the simulated galaxies, causing them to deviate further from pure exponential profiles and giving larger overall sizes.   This is presumably because not enough gas is penetrating down to small radii to fuel star-formation there. 
This suggests the need for one or more physical processes to transport angular momentum of disk gas outward in the simulations, in order to reduce the sizes of the simulated disks.

\subsection{The stellar surface density profiles} \label{sec:3.3}

\begin{figure*}[htb]
  \centering
\includegraphics[width=0.4\textwidth]{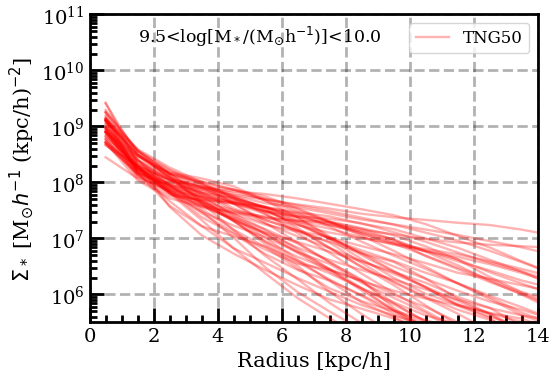}
\includegraphics[width=0.4\textwidth]{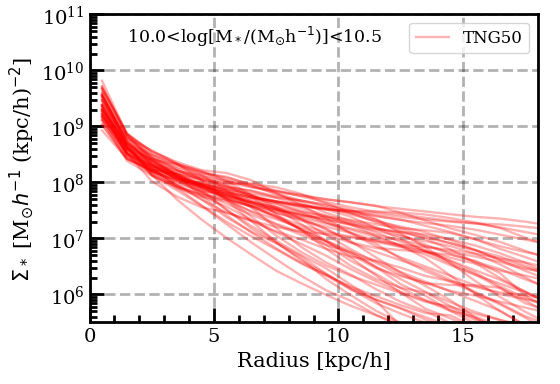}
\includegraphics[width=0.4\textwidth]{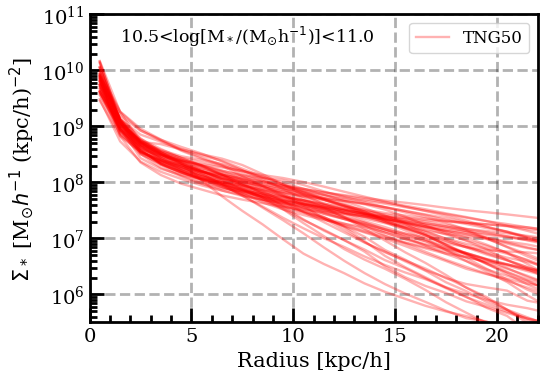}
\includegraphics[width=0.4\textwidth]{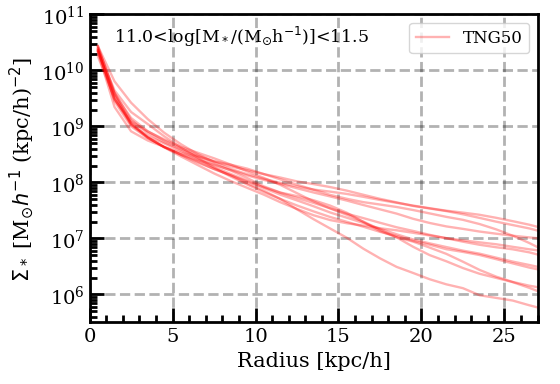}
  \caption{The stellar surface mass density profiles for TNG50 galaxies. As above, we show the result for the four stellar mass bins, as denoted in each panel. }
  \label{fig:7}
\end{figure*}

\begin{figure*}[htb]
  \centering
\includegraphics[width=0.9\textwidth]{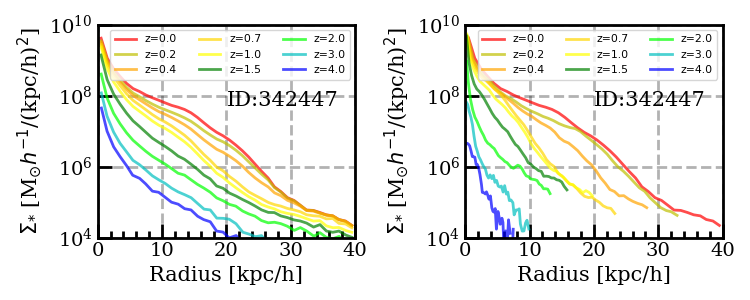}
\includegraphics[width=0.9\textwidth]{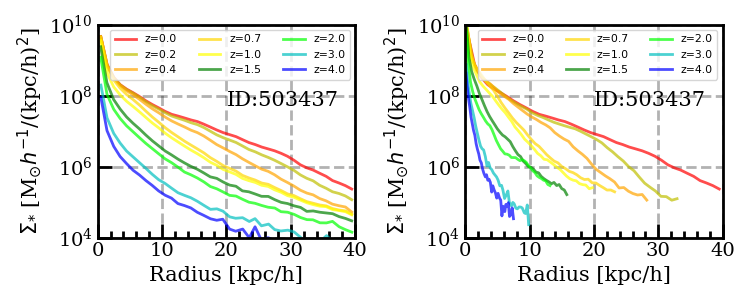}
  \caption{ Illustration of the radial migration of stellar particles with two TNG50 individual galaxies: ID-342447 (top panels) and ID-503437 (bottom panels). Left panels: The radial distribution of the $\Sigma_*$ {\it at the current epoch} ($z=0.0$) for those stellar particles that formed prior to a set of different redshifts.  Right panels: The evolution of $\Sigma_*$ for the individual galaxies by tracing their main progenitors in the merging tree at different snapshots.  Therefore, by comparing of left and right panels, one can imagine that the differences are due to the radial migration of stellar particles and mergers. }
  \label{fig:8}
\end{figure*}

\begin{figure*}[htb]
  \centering
\includegraphics[width=0.4\textwidth]{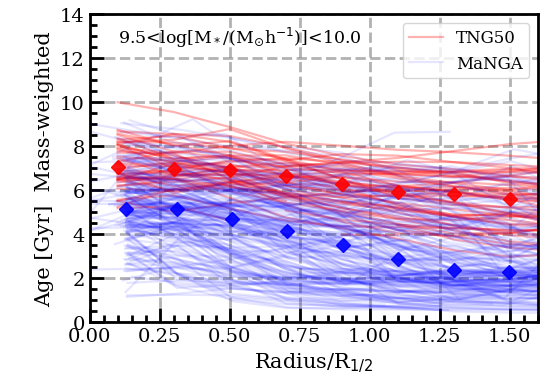}
\includegraphics[width=0.4\textwidth]{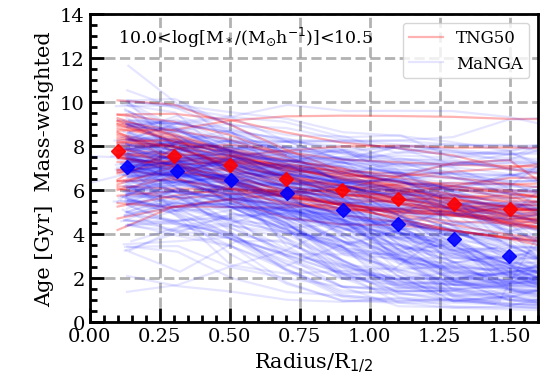}
\includegraphics[width=0.4\textwidth]{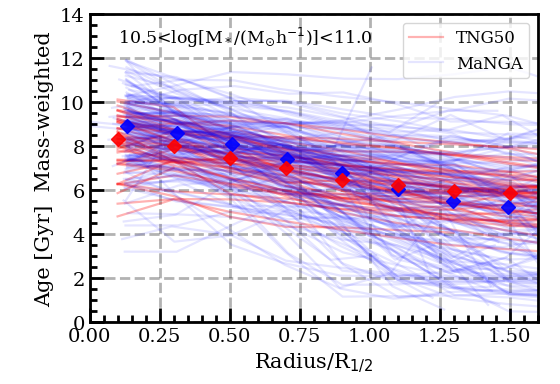}
\includegraphics[width=0.4\textwidth]{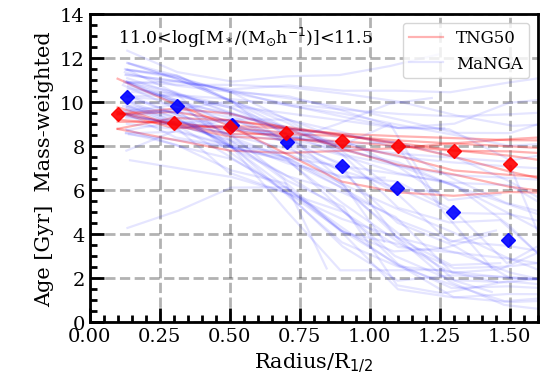}
  \caption{The mass-weighted age profiles for TNG50 galaxies (red lines) and MaNGA galaxies (blue lines). The red and blue diamonds show the median age profiles for TNG50 and MaNGA, respectively. As above, we show the result in the four stellar mass bins.  }
  \label{fig:9}
\end{figure*}

In the previous subsection, we showed that the $\Sigma_{\rm SFR}(r)$ profiles of TNG50 galaxies show stronger deviations from pure exponential functions than do real MaNGA galaxies.  In this subsection, we turn to examine the resulting profiles of stellar surface density ($\Sigma_*(r)$) for the TNG50 galaxies.  

Figure \ref{fig:7} shows the $\Sigma_*(r)$ profiles of the TNG50 galaxies for the four stellar mass bins.  Interestingly, we find from simple visual inspection that the $\Sigma_*(r)$ profiles typically show very good exponential disks with a S{\'e}rsic core (or bulge) in the galactic center.  This appears to be in good agreement with the observations on the stellar distribution of disk galaxies \citep[e.g.][]{Kent-85, Weiner-01, Pohlen-06, Meert-13}.  It is interesting that the $\Sigma_{\rm SFR}(r)$ profiles do not show good exponential form but the $\Sigma_*(r)$ profiles do, since the $\Sigma_*$ might be expected to be the time-integration of $\Sigma_{\rm SFR}$, if there is no radial migration of stars\footnote{Significant mergers have already being excluded in the sample selection of TNG50 galaxies since redshift of $0.4$ (see Section \ref{sec:2}).}.   Indeed, \cite{Lilly-16} have shown that good exponential stellar disks with a superposed bulge, as well as realistic radial gradients of sSFR with radius, can be produced by straightforward superposition of (pure) exponential $\Sigma_{\rm SFR}(r)$, assuming that the evolution of the overall SFR and the exponential scalelength of $\Sigma_{\rm SFR}(r)$ follow the relations indicated from observations of real galaxies at different redshifts.

To explore this question further, we examine the importance of radial migration of stars in two individual galaxies (TNG50 ID: 342447 and 503437).  These two galaxies are not special, and can be treated as a representative of the TNG50 galaxies, at least on the question we considered here.   The radial redistribution of stars can be seen by comparing the radial distribution of (all) stars when they are actually formed with the radial distribution of the same set of stars at the current epoch. 
The left panels of Figure \ref{fig:8} show the $\Sigma_*(r)$ profiles {\it at the current epoch} ($z=0.0$) for the stellar particles in these two galaxies that were formed prior to a set of different redshifts.  The right panels of Figure \ref{fig:8} show the evolution of $\Sigma_*(r)$ for these two galaxies {\it at different epochs}, which is obtained by tracking their main progenitors through the simulation.   If there was no redistribution of stars after they formed, then these two representations would obviously be identical.

As can be seen in the right panels of Figure \ref{fig:8}, the size of the main progenitors gradually increases with time for both galaxies. This is consistent with the significant size evolution seen in observations \citep[e.g.][]{Buitrago-08, Newman-12, Mosleh-12, van-der-Wel-14, Genel-18}.   Specifically, the size of the stellar disk for the main progenitors is very small before $z=1.5$.  However, the distribution at the current epoch of those stars that formed before $z=1.5$ show a much more extended distribution than when they were actually formed.   Some of this could well be due to mergers of the galaxies, but it is noticeable that the effect is seen even at redshifts below $z=0.4$, when significant mergers have been excluded.   This comparison indicates that there is a strong radial migration of stars in the evolution of the stellar disk in the simulated TNG50 galaxies.   

It is then interesting to understand why the radial migration of stars can result in an exponential stellar disk. As mentioned in the Introduction (Section \ref{sec:1}), \cite{Elmegreen-13} proposed that stellar scattering by encountering with massive clumps can redistribute stars and  produce exponential profiles in a self-gravitating system \citep{Wu-20}, which may be due to the fact that the exponential stellar disk has the maximum entropy state for the distribution of specific angular momentum of stars \citep{Herpich-17}.  

However, it is not clear whether the scattering of stars is as important in real galaxies as in the simulations.  
One diagnostic of radial redistribution is the radial profile of stellar age, which is likely to be flattened by radial redistribution of stars.  Therefore, we examine the profiles of mass-weighted stellar age for TNG50 galaxies (red lines) and as estimated for MaNGA galaxies (blue lines) of the four stellar mass bins in Figure \ref{fig:9}.  In each mass bin, we also show the median age profiles of the TNG50 sample and MaNGA sample in
the red and blue diamonds, respectively.  
The stellar age profiles are taken from \cite{Wang-18}, obtained from the output of {\tt STARLIGHT} code \citep{Cid-Fernandes-05}.   In the {\tt STARLIGHT} fitting, the templates are 45 single stellar populations taken from \cite{Bruzual-03} with assuming a \cite{Chabrier-03} IMF and a \cite{Cardelli-89} stellar extinction law.  Although the stellar age of MaNGA obtained from the {\tt STARLIGHT} fitting is not model-independent, \cite{Wang-18} pointed out that the gradients in stellar age obtained by {\tt STARLIGHT} for MaNGA galaxies show good consistency with those of the output from {\tt pPXF} and {\tt FIREFLY} \citep{Zheng-17, Li-18, Goddard-17}. 

As shown in Figure \ref{fig:9}, the overall negative gradients in stellar age profiles are in good general agreement with an inside-out growth scenario for disk galaxies \citep[e.g.][]{Perez-13, Li-15, Ibarra-Medel-16, Wang-18}, for both TNG50 and MaNGA.  However, the TNG50 galaxies show significantly flatter age gradients than do real MaNGA galaxies, across the full range of stellar mass. This supports our previous suspicion that the radial migration of stars in TNG50 galaxies is much larger than in real galaxies, and that the good exponential distributions of stellar mass that are seen in TNG50 galaxies (despite their less exponential $\Sigma_{\rm SFR}(r)$ profiles) may plausibly reflect this.

\section{Gas inflow, outflow and star formation in the TNG50 disks} \label{sec:4}

In previous section, we found that simulated TNG50 galaxies typically show larger sizes and larger deviations from pure exponential $\Sigma_{\rm SFR}$ profiles than do real MaNGA galaxies.  The good exponential profiles of stellar mass density $\Sigma_*$ in TNG50 galaxies are likely to be the result of large scale radial migration of stars by stellar scattering, which may be not realistic because it produces too flat age gradients compared to observations. In this section, we focus on the gas accretion onto gas disks, in order to examine to what extent galaxies in the simulation behave as gas-regulator systems \citep[e.g.][]{Lilly-13}. 

\subsection{The definition of inflow and outflow rates} \label{sec:4.1}

\begin{figure*}[htb]
  \centering
\includegraphics[width=0.4\textwidth]{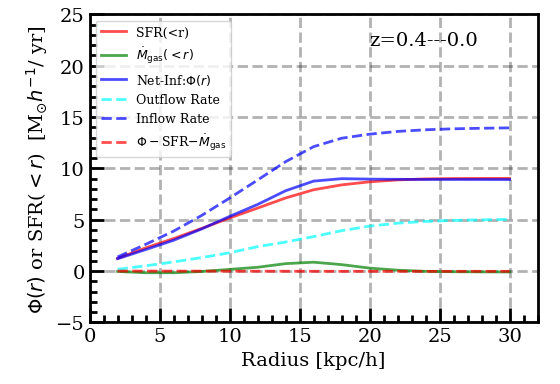}
\includegraphics[width=0.4\textwidth]{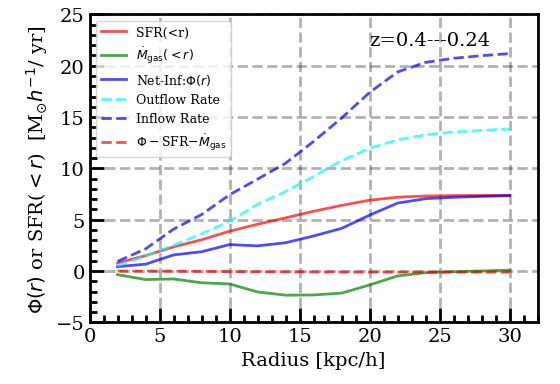}
\includegraphics[width=0.4\textwidth]{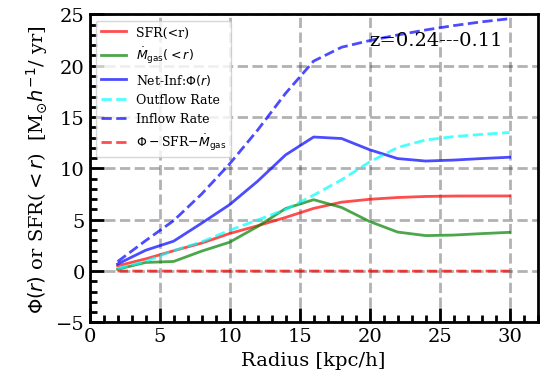}
\includegraphics[width=0.4\textwidth]{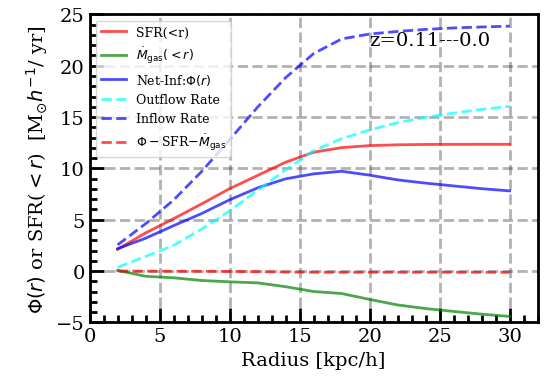}
  \caption{  Illustration of the definition of net inflow rate with one individual TNG50 galaxy (ID:342447).  
  Top left panel: the red line shows the cumulative SFR as a function of radius, i.e. SFR($<r$). The blue dashed line shows the inflow rate as a function of radius, the cyan dashed line shows the outflow rate as a function of radius, and the blue solid line shows the net inflow rate as a function of radius.  The green line shows the change rate of cumulative gas mass as a function of radius.  We can check that the net inflow rate within a given $r$ equals the SFR($<r$) plus the $\dot{M}_{\rm gas}(<r)$, which is shown in the red dashed line (net inflow rate minus SFR($<r$) and  $\dot{M}_{\rm gas}(<r)$). We note that all these quantities are measured between the two snapshots of $z=0.0$ and $z=0.4$. 
  The other three panels:  The same as the top left panel, but all these quantities are measured between two snapshots of shorter time intervals, as denoted in each panel.  
}
  \label{fig:10}
\end{figure*}

There are usually two approaches to computing the inflow or outflow rate: 1) deriving instantaneous fluxes of mass that are based on gas velocities 
at a given time, i.e. in one snapshot of the simulation \citep[e.g.][]{Ocvirk-08, Nelson-19} and 2) deriving mass fluxes across a boundary 
by tracking the movement of gas particles between two different snapshots \citep[e.g.][]{Nelson-13} which necessarily averages the inflow/outflow over the time interval between the snapshots.  In this work, we adopt the latter approach, since our main focus is the gas inflow and outflow on relatively long timescale, upto a few Gyr.  

We compute the inflow rate and outflow rate of our sample of TNG50 galaxies in the following way.  For a given galaxy in the simulation, we first extract the particle information (at different snapshots) for its main progenitor from the merging tree.  We then define the average (between any two snapshots) inflow and outflow rate across any given boundary.  Specifically, for two snapshots S1 (at redshift $z1$) and S2 (at redshift $z2$, and $z1>z2$), the inflowing gas particles within the time interval are defined as those gas particles, plus those stellar particles that are formed between the two snapshots, that appear within the given volume in S2 but are not there in S1.  In the same way, the outflowing gas particles are defined as the gas particles that are in a given volume at S1 but not in the corresponding volume at S2.  The inflow rate (or outflow rate) is then the sum of all inflowing (or outflowing) gas particles divided by the time interval between the two snapshots. 
The tracing of gas mass by their particle IDs, as adopted here, ignores the change of mass of individual particles due to mass exchange between them, that is occurs during the ``refinement'' and ``de-refinement'' procedures used in the TNG simulation to maintain the average mass of gas particles within a factor of two of a given value. We have checked that the net effect of these changes of individual particle masses is infact negligible with respect to the total gas mass, and may, as here, be ignored, at least for the purpose of this analysis.

The boundary of the volume is defined in the following way.  For a given TNG50 galaxy, we first convert the coordinates of all the particles in this galaxy into cylindrical coordinates, with the zero point as the galactic center and the $z$-axis as the direction of the disk axis, calculated as above.  
We calculate the inflow and outflow rates using the above approach across a cylindrical boundary with $|z| = 5$ kpc and a variable radius  $r$.  By increasing the radius $r$ we can therefore obtain the inflow (or outflow) rate of the disk as a function of galactocentric radius. 

For illustration, Figure \ref{fig:10} shows the inflow rate and outflow rate as a function of radius for one particular galaxy (TNG50 ID: 342447) between different pairs of snapshots, as denoted in each panel.  In each panel, we show as a blue solid line the net inflow rate as a function of radius, denoted as $\Phi(r)$ and defined as the inflow rate minus the outflow rate through the cylindrical surface of radius $r$ described above.  For comparison, we also show the cumulative SFR($<r$) as a function of radius (as the red solid lines), computed as the time-averaged (between the two snapshots) SFR within the radius $r$.  In addition, we also show as a green solid line average the rate of change of the total gas mass within the radius $r$, denoted as $\dot{M}_{\rm gas}(<r)$.  
Although we set a fixed cylindrical boundary of $|z| = 5$ kpc, the resulting net inflow rate is not sensitive to this choice. We have examined that the net inflow rates generally only change by a few percent when $|z|$ is varied from 3 kpc and 10 kpc, and the main conclusion from this section therefore does not change.

The conservation of mass then implies that these quantities should be related by.
\begin{equation} \label{eq:2}
 \dot{M}_{\rm gas}(<r) = \Phi(r) - {\rm SFR}(<r). 
\end{equation}
We verify this equation by showing the $\Phi(r) - {\rm SFR}(<r)-\dot{M}_{\rm gas}(<r)$ as a function of radius as the red dashed line in each panel of Figure \ref{fig:10}.  

It should be noted that our estimate of the net inflow rate will be affected by stellar mass loss.  Although in TNG50, the effects of stellar mass loss are incorporated by increasing the mass of nearby gas particles, the refinement/de-refinement processes ensure that the average mass of gas particles stays approximately constant, and therefore the mass injection from gas return from stars eventually leads (effectively) to the creation of new gas particles that, in our approach, will lead to an over-estimate of the mass inflow rate\footnote{This is also the reason why we can verify the mass conservation in Equation \ref{eq:2}.}.   Apart from this effect of stellar mass return, the splitting or merging of gas particles during refinement/de-refinement should not affect the net inflow because the perturbations of the inflow and outflow, which will arise from splitting and merging respectively, should cancel out. It will be shown in Section \ref{sec:4.2} that this gas return component is small compared with the true inflow of pre-existing gas particles.   

It should be noted that the SFR we considered throughout this work is the instantaneous formation rate of newly formed stars, rather than the formation rate of long-lived stars, sometimes, known as the reduced SFR.  Therefore, comparison between our defined $\Phi$ and SFR is appropriate to examine whether the TNG galaxies can be treated as gas-regulator systems. This also means that Equation \ref{eq:2} is satisfied.    To eliminate this effect,  one could subtract the gas return rate, which is $\sim$30\% of SFR for the TNG simulation  \citep{Pillepich-18}, from our defined $\Phi$ (and the SFR). This does not change our main conclusions. 

In the top left panel of Figure \ref{fig:10}, all the quantities are computed between $0.0<z<0.4$. However, for the other three panels, the quantities are computed for three shorter intervals $0.24<z<0.4$, $0.11<z<0.24$, and $0.0<z<0.11$, respectively.  Interestingly, the inflow (and outflow) rate is higher for a factor of $\sim$1.6 when measured in the narrower time intervals, while the net inflow rate calculated between $0.0<z<0.4$ appears to be comparable to the ones calculated within the three narrower time intervals (of order 1.5 Gyr).  
This suggests gas recycling is very important in the simulated galaxies, in the sense that it implies that a significant number of gas particles are being registered as entering the galaxy in one of the shorter time intervals and then registered as leaving in another interval and are therefore not counted as having crossed the boundary within the longer time interval. We have examined this for the galaxy (ID: 342447) during the epochs after redshift of 0.4.  About 40\% of inflowing gas particles, defined as those appearing between a given pair of snapshots (with mean $\Delta t \sim$ 170 Myr), have then left the galaxy before the next snapshot, and are therefore classified as outflowing particles between this next pair of snapshots. 

To avoid this issue, in this work, we only use the {\it net} inflow rate computed over the whole interval $0.0<z<0.4$ in the following analysis.  

\subsection{Gas accretion supports star formation in galaxies} \label{sec:4.2}

\begin{figure*}[htb]
  \centering
\includegraphics[width=0.4\textwidth]{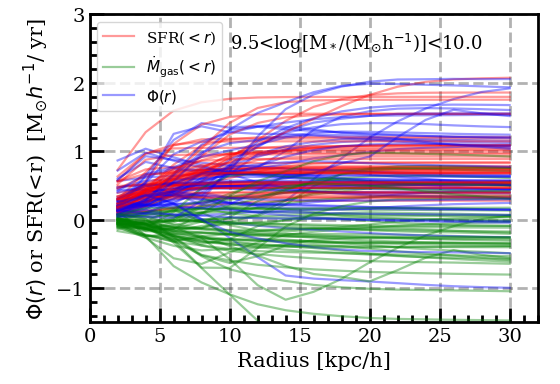}
\includegraphics[width=0.4\textwidth]{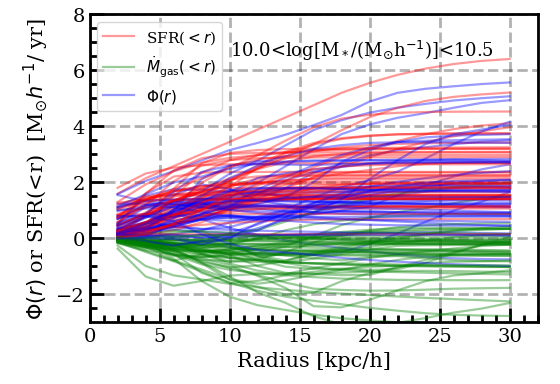}
\includegraphics[width=0.4\textwidth]{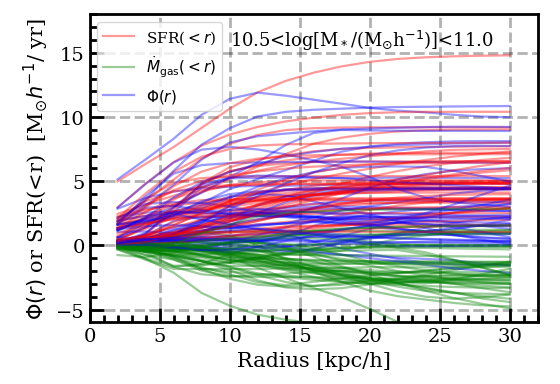}
\includegraphics[width=0.4\textwidth]{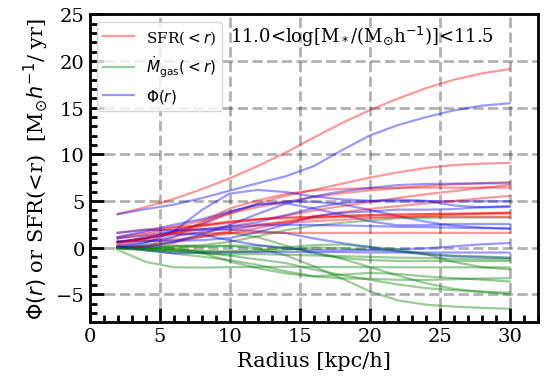}
  \caption{The cumulative SFR($<r$), net inflow rate and $\dot{M}_{\rm gas}(<r)$ as a function of radius for the TNG50 galaxies.   As above, galaxies are separated into the four stellar mass bins in showing the result.  }
  \label{fig:11}
\end{figure*}

\begin{figure*}[htb]
  \centering
\includegraphics[width=0.4\textwidth]{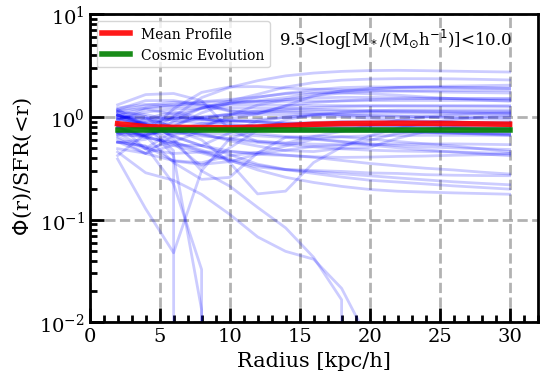}
\includegraphics[width=0.4\textwidth]{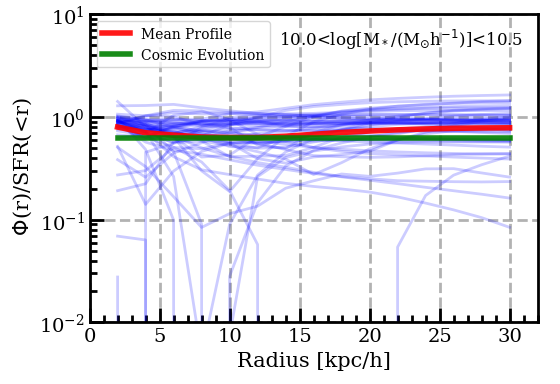}
\includegraphics[width=0.4\textwidth]{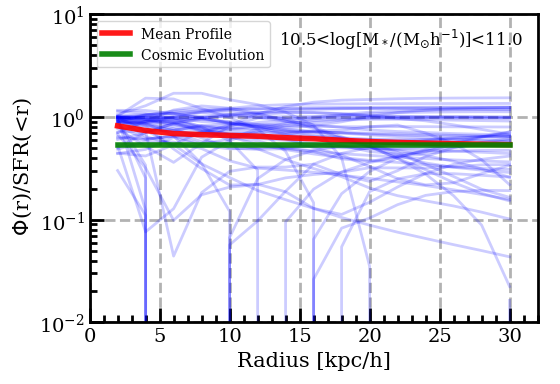}
\includegraphics[width=0.4\textwidth]{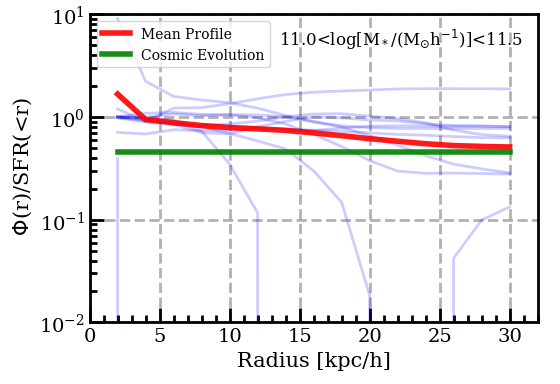}
  \caption{ The ratio of net inflow rate to SFR($<r$) as a function of radius for the TNG50 galaxies of the four stellar mass bins. In each panel, the red line shows the mean profile of the TNG50 galaxies of the corresponding stellar mass bin. The green horizontal line shows the value of $\Phi/$SFR expected from the cosmic evolution of SFMS (see text).  }
  \label{fig:12}
\end{figure*}

Comparison of the relative sizes of the net inflow rate, the star-formation rate, and the rate of change of gas mass provides a basic check of the gas regulator picture which requires that the rate of change of gas mass be smaller than the other two.   
Figure \ref{fig:11} shows the net inflow rate $\Phi(r)$, the cumulative SFR($<r$), and the $\dot{M}_{\rm gas}(<r)$ as a function of radius for individual TNG50 galaxies for each of the four stellar mass bins.  
The $\dot{M}_{\rm gas}(<r)$ of TNG50 galaxies are typically negative across the full range of stellar mass and over the full range of radii.  This indicates an overall reduction of gas mass in the disk over the last 4.4 Gyr.   In the gas regulator picture, such a reduction is required in order to be consistent with the cosmic evolution of individual galaxies implied by the evolution of the sSFR of the SFMS
\citep[e.g.][]{Noeske-07, Daddi-07, Peng-10, Stark-13} provided that the star-formation efficiency is more or less constant. 

As a whole, both the $\Phi(r)$ and the SFR$(<r)$ appear to be larger than the absolute value of $\dot{M}_{\rm gas}(<r)$.
Specifically, we calculate the mean value of $\Phi/|\dot{M}_{\rm gas}|$ at $r=$30 kpc for the four stellar mass bins, and find these to be 5.3, 4.2 1.7 and 1.5 respectively with increasing stellar mass.  This indicates that the gas fuel for star formation in TNG50 galaxies is primarily provided by gas accretion, rather than by consumption of a pre-existing gas reservoir within the galaxies, at least for galaxies with $\log[M_*/({\rm M}_{\odot}h^{-1})]<10.5$. For galaxies at the higher masses, the change of gas mass $\dot{M}_{\rm gas}$ becomes more significant. We suggest that this is due to the overall reduction of SFR for massive star-forming galaxies (see later Figure \ref{fig:12}). 

To illustrate this more clearly, we directly show the {\it ratio} of $\Phi(r)$ and SFR($<r$) as a function of radius for individual TNG50 galaxies, in Figure \ref{fig:12}.  For each stellar mass bin, we compute the mean value of $\Phi(r)$/SFR($<r$) as the red solid line.  For comparison, we also show the values of $\Phi$/SFR that are expected from the cosmic evolution of the SFMS in the observation for the corresponding stellar mass bins, in green horizontal lines.  

This predicted $\Phi$/SFR in the following way.  Motivated from the observations \citep[e.g.][]{Daddi-07, Elbaz-11, Stark-13}, we assume that the evolution of specific SFR, defined as SFR/$M_*$, follows the formula \citep{Lilly-16, Wang-22}:
\begin{equation} \label{eq:3}
  {\rm sSFR}(M_*,z) = \frac{0.07}{1-R}\times  (\frac{M_*}{3\times 10^{10}M_{\odot}})^{-0.2}\times (1+z)^2 \ {\rm Gyr^{-1}}, 
\end{equation}
where $R$ is the fraction of mass formed in new stars that is subsequently returned to the interstellar medium through winds and supernova explosion.  We take $R=0.4$ for the \cite{Chabrier-03} IMF \citep[see][]{Vincenzo-16}. Based on Equation \ref{eq:3}, we can then derive the star formation histories for individual galaxies of any stellar mass at the current epoch.  We adopt a typical gas depletion timescale of $\tau_{\rm gas}=$5.4 Gyr (including both atomic and molecular gas) taken from \cite{Saintonge-17}, which is nearly independent of galaxy mass. Based on the above assumptions, we can then obtain the predicted $\Phi$/SFR between redshift of 0.0 and 0.4 for galaxies of any given stellar mass:  
\begin{equation} \label{eq:4}
 \Phi/{\rm \langle SFR \rangle}  = 1 - \frac{\dot{M}_{\rm gas}}{\rm \langle SFR \rangle} 
             =  1- \frac{\Delta \rm SFR}{\Delta t} \times  \tau_{\rm gas}/{\rm \langle SFR \rangle},
\end{equation}
where ${\rm \langle SFR \rangle}$ is the mean SFR within the time interval.  

As can be seen in Figure \ref{fig:12}, the radial profiles of $\Phi(r)$/SFR($<r$) are overall flat for individual TNG50 galaxies across the full range of stellar mass.   The mean values of $\Phi(r)$/SFR($<r$) are slightly less than unity, as expected from the cosmic evolution of SFR in galaxies.   
The key requirement of the gas regulator picture is that the rate of change in gas mass is small compared with the SFR and net inflow of the system.  This is clearly satisfied for TNG50 galaxies, at least for galaxies with $\log[M_*/({\rm M}_{\odot}h^{-1})]<10.5$. Allowing for the cosmic reduction of SFR for massive star-forming galaxies, TNG50 galaxies can therefore be treated as gas-regulator systems in which gas continuously flows through the system, rather than simple ``close-box'' or ``leaky-box''  scenarios in which a gas reservoir is ``consumed" \citep{Lilly-13}.  The gas reservoir, which must represent a balance between inflow, star-formation and wind-driven outflow, adjusts so as to maintain the SFR close to that required to consume the inflowing gas  \citep{Schaye-10, Bouche-10, Dave-11, Lilly-13, Belfiore-19}. 

The notion of ``gas-regulator'' was proposed by \cite{Lilly-13} followed the work of \cite{Dave-12}.  This simple idea of galaxy formation appears to explain a large range of both observational and simulated data. Specifically, \cite{Dave-12} provided an analytic formalism to describe the evolution of stellar mass, gas mass and metallicity of galaxies, assuming an equilibrium state that the mass of the gas reservoir is constant with time. This scenario is known as the ``bathtub'' model \citep{Bouche-10, Dave-12}. However, releasing this restriction of zero change in gas mass, \cite{Lilly-13} proposed that the SFR in galaxies is regulated by the instantaneous gas mass continually adjusting to the inflow rate. This is known as the ``gas-regulator'' model.  In the gas-regulator framework, the SFR emerges naturally as a second parameter in the mass-metallicity relation \cite{Lilly-13}, and therefore provides a natural explanation for the claimed universal (epoch independent) mass–metallicity-SFR relation \citep[known as the fundamental metallicity relation;][]{Mannucci-10, Nakajima-12, Dayal-13, Salim-14, Cresci-19, Curti-20}.

Further, in the same framework of the gas-regulator picture, \cite{Wang-19} found that the dispersion of $\Sigma_{\rm SFR}$ across the galaxy populations can be quantitatively explained by the response of gas regulator systems to temporal variations of gas inflow rate on timescales of a few Gyr.  For a given variation of inflow, the amplitude of the response of the regulator depends on the gas consumption timescale of the system.   Interestingly, this time-variation explanation of the dispersion in  $\Sigma_{\rm SFR}$ is further supported by a direct analysis of the temporal changes in the SFR as indicated by comparisons of the H$\alpha$ emission and H$\delta$ absorption lines \citep{Wang-20b}.   

Consistent with this, by exploring the time-variability of star formation histories for galaxies taken from hydrodynamical simulations and semi-analytic models, \cite{Iyer-20} found that the dark matter accretion histories of galaxies are in general coherent with the in-situ star formation on timescales $>$3 Gyr \citep[see also][]{Tacchella-20}.  More recently, \cite{Wang-21} further developed the gas-regulator model to understand the correlations between SFR, gas mass, and metallicity, as well as their time-variability, on different spatial scales.  \cite{Wang-21} found that the negative correlation between SFR and metallicity on galactic scale is the result of a time-varying inflow rate, while the positive correlation between these two quantities on 100 pc scale within individual galaxies is the result of a time-varying star formation efficiency as gas orbits around the galaxy \citep{Kreckel-18, Kruijssen-19, Chevance-20}. 

\section{The evolution of angular momentum of inflowing gas particles} \label{sec:5}

\begin{figure*}[htb]
  \centering
\includegraphics[width=0.32\textwidth]{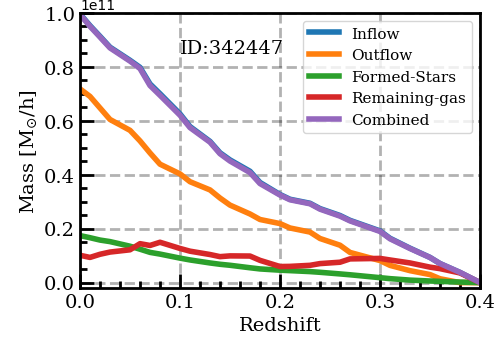}
\includegraphics[width=0.32\textwidth]{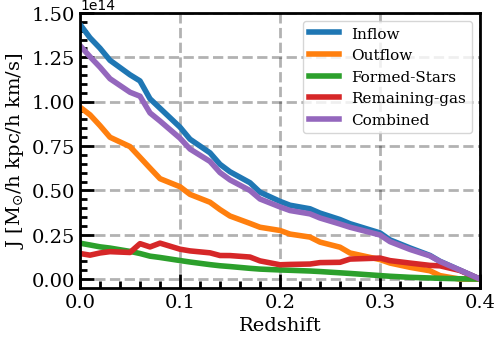}
\includegraphics[width=0.32\textwidth]{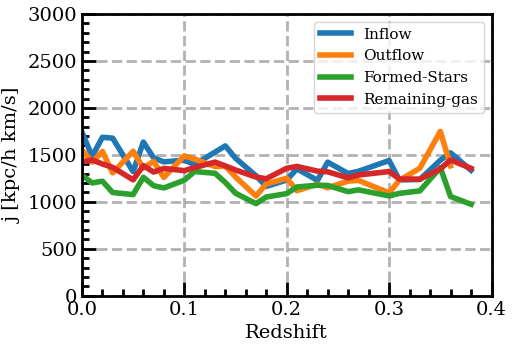}

\includegraphics[width=0.32\textwidth]{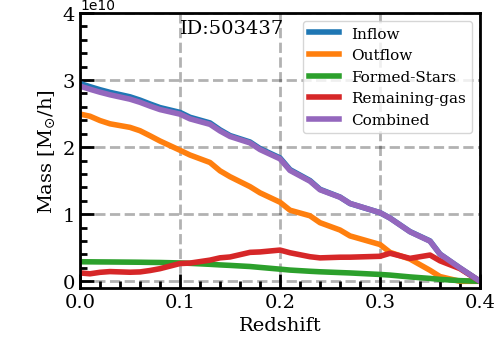}
\includegraphics[width=0.32\textwidth]{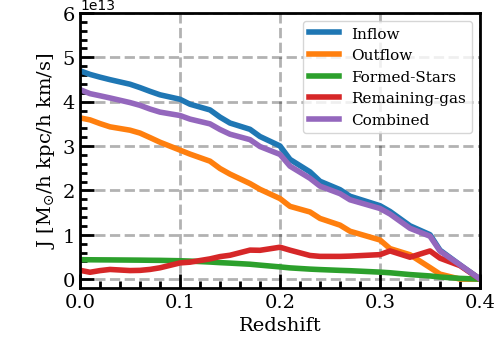}
\includegraphics[width=0.32\textwidth]{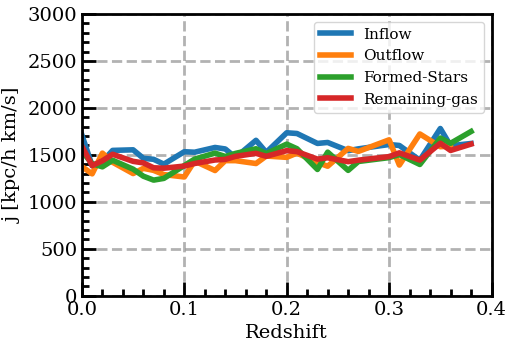}
  \caption{  The angular momentum evolution of the inflowing gas particles since redshift of 0.4 for two individual galaxies: ID-342447 (top panels) and ID-503437 (bottom panels).  The inflowing gas particles can later be either outflowing gas particles, stellar particles or remaining gas particles (see text).  We therefore show the cumulative mass of inflowing gas particles, as well as the mass of three different components (the cumulative mass of outflowing particles, the instantaneous mass of  stellar and gas particles) in the left column of panels.  We can check that the combination of the three components equals to the cumulative mass of inflowing gas particles at any epoch for consistency check. 
  The middle column of panels show the cumulative angular momentum for the inflowing gas particles, as well as the three different components (the cumulative angular momentum for the outflowing gas particles, and the angular momentum of instantaneous stellar and gas particles).  The right column of panels show the evolution of mean specific angular momentum of the inflowing gas, outflowing gas, remaining gas and the newly formed stellar particles (defined at the corresponding epochs rather than the cumulative particles). }
  \label{fig:13}
\end{figure*}

\begin{figure*}[htb]
  \centering
\includegraphics[width=0.4\textwidth]{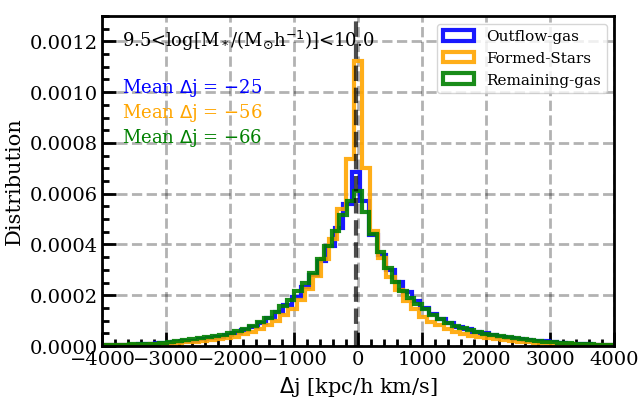}
\includegraphics[width=0.4\textwidth]{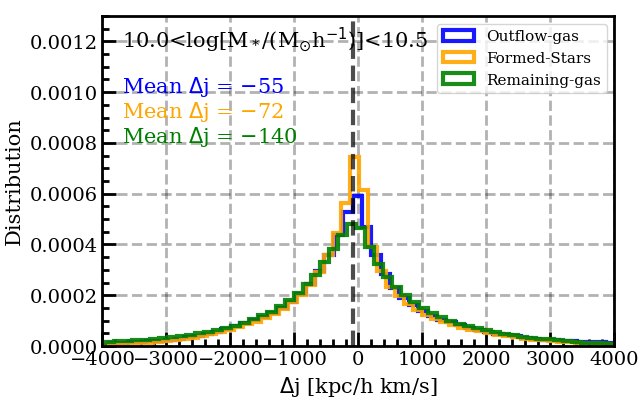}
\includegraphics[width=0.4\textwidth]{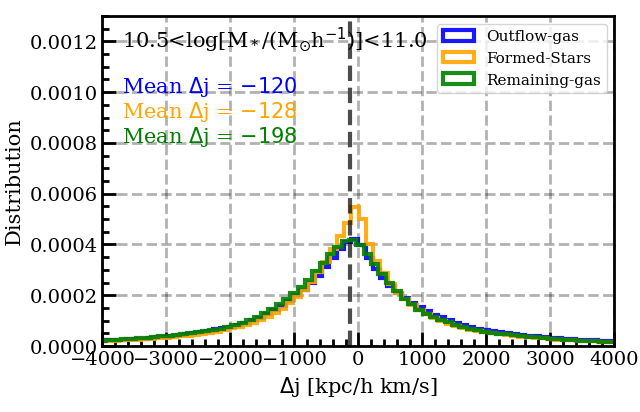}
\includegraphics[width=0.4\textwidth]{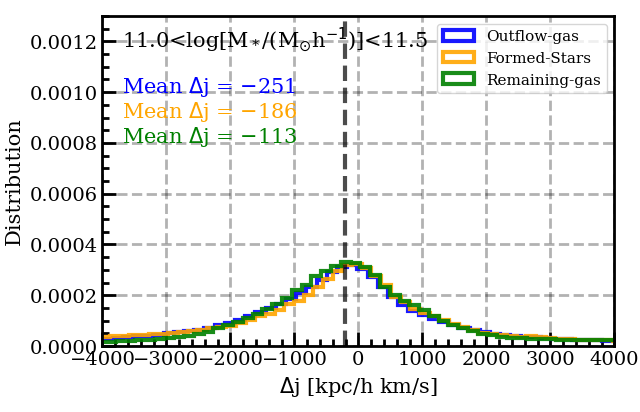}
  \caption{  The mass-weighted distribution of the change of specific angular momentum, $\Delta$j (see the detailed definition in text), for individual inflowing gas particles (being either outflowing gas particles, stellar particles or remaining gas particles).  We calculate the $\Delta$j for all the sample galaxies of TNG50 by tracing the inflowing particles from $z=0.4$ to $z=0.2$ (to save computing time). As above, the results are shown for the four stellar mass bins.  In each panel, we indicate the mass-weighted average $\Delta j$ for the three cases with the corresponding colors, and show the mass-weighted average $\Delta$j for all of the inflowing particles in the vertical dashed line in each panel. } 
  \label{fig:14}
\end{figure*}

In the previous section, we established that, as expected, the TNG50 simulated galaxies can be treated as gas regulator systems in which the inflowing gas is the key to driving the cosmic evolution of star formation in galaxies \citep[see also][]{Iyer-20}.  It is then interesting to examine the histories of the inflowing gas particles, especially the evolution of their angular momenta, in order to better understand the reason for the somewhat large disk sizes and non-exponential $\Sigma_{\rm SFR}$ profiles that are seen in TNG50 galaxies (see Section \ref{sec:3.1}).  

Therefore, we examine the angular momentum ($J$) of the gas particles when they are first accreted through our boundary, and any change of $J$ afterwards.   Figure \ref{fig:13} illustrates the evolution of inflowing gas particles for two individual galaxies (TNG50 ID: 342447 and 503437) taken as examples.  For individual galaxies, we record all the gas particles that are identified as inflowing, beginning at $z \sim 0.4$, and then follow these particles in all later snapshots down to $z \sim 0.0$.  For any two adjacent snapshots, the inflowing particles are defined in the same way as in Section \ref{sec:4}, with the boundary being set as $|z|<$ 5 kpc and $r<10$ kpc in cylindrical coordinates centered on the galactic center in each of the snapshot.   We adopt this same boundary for all the individual galaxies in the analysis of this section.  

These identified inflowing gas particles can then, at some later time, either leave the volume as outflowing gas particles, be converted into stellar particles through star-formation, or remain as gas particles within the volume.  We stress that in this section, we only consider those outflowing, stellar particles or gas particles that entered the boundary as inflowing gas particles at times after $z \sim 0.4$. 

We consider the angular momentum for the different type of particles in the following way.  The $J$ of inflowing particles are computed only when the gas particles first appear within the defined boundary of the galaxies.   In the subsequent evolution, the $J$ of outflowing gas particles are computed using the last snapshot in which they were within the defined boundary.  We may therefore conceivably underestimate the $J$ of such particles, if they acquire additional angular momentum immediately before they pass through the boundary. This is a technical limitation that we cannot easily track these particles after they have left the simulated galaxies (without loading the full set of particles of a given snapshot).  The $J$ of the stellar particles are computed in the snapshot in which they first become stellar particles through star-formation.  

In the left panels of Figure \ref{fig:13}, we show the evolution of the cumulative mass of inflowing gas particles as the blue solid line, and the three other components in orange (cumulative mass of outflowing gas), green (instantaneous mass of stellar particles) and red (instantaneous mass of remaining gas particles) solid lines.  The first three of these steadily increase with time.   As required, the cumulative mass of the inflowing gas particles equals at all times the cumulative mass of the outflow plus the instantaneous mass of the stars and remaining gas. 

We show the cumulative $J$ for inflowing gas particles as a function of redshift, the cumulative $J$ of the outflowing particles, and the instantaneous $J$ of the stellar and gas particles, in the middle panels of Figure \ref{fig:13}, as denoted by different colors.  Note that the cumulative inflow is much larger than the mass of stars, and thus of the mass return from stars, so most of the ``inflow'' is real inflow and not mass-return from stars.  Additionally we show the sum of the three last of these (outflow, stars and remaining gas) as a purple line.
Interestingly, we find this combined angular momentum to be slightly less than (though by $<$10\%)  the $J$ of the inflowing particles.  On the one hand, this indicates a small loss of angular momentum after particles appear as inflow.
This small loss may due to the underestimation of angular momentum for the outflowing particles as discussed above. 
However, the fraction of angular momentum is very small compared to their original $J$, which further suggests that the processes for transporting angular momentum of gas particles within the simulation are not very effective.

In the right panels of Figure \ref{fig:13}, we show the evolution of the mean (mass-weighted) specific angular momentum, defined as $j=J/M$, for the inflowing particles, outflowing  particles, the newly-formed stellar particles and the remaining gas particles.   It should be noted that, unlike the left and middle panels of Figure \ref{fig:13},  the $j$ of the inflowing, outflowing and newly-formed stellar particles in the right panels of Figure \ref{fig:13} are not cumulative quantities, and represent the $j$ of the particles that are identified (as inflow, outflow, newly formed star particle, or within the gas reservoir) at each epoch.

We find the $j$ of the inflowing particles does not show significant evolution since redshift of 0.4, for these two particular galaxies. Actually, we have examined the evolution of $j$ of the inflowing particles for the sample galaxies, and find that this is true for the majority of TNG50 galaxies. 
This is also the case for the outflowing, newly-formed stellar particles and remaining gas particles.  

It appears that the inflowing particles show slightly higher $j$ than the other three kinds of particles, while the newly-formed stellar particles show slightly lower or comparable $j$ than the other three kinds of particles.  This suggests that the initial angular momentum of inflowing particles may be weakly related to their fates, in the sense that, inflowing gas particles with lower $j$ tend to have a larger chance to be converted into a star particle, rather than being ejected as an outflowing particle later.  This might be thought to be due to the fact that gas particles of lower $j$ may fall into inner regions of galaxies, where the star formation efficiency is relatively high.

We then explore the evolution of $j$ for individual inflowing particles, by comparing the $j$ of individual particles at the time of their being accreted to the $j$ of the same particles when they are later converted into stars, when they left the system as outflowing particles, or at the last snapshot we considered if they were still gas particles. We define the change of $j$ for individual particles between entry and their eventual fate as $\Delta j$.  We track the inflowing gas particles for all our sample galaxies of TNG50 from redshift of 0.4 to 0.2\footnote{We do not track the particles to redshift of 0.0 only to save computing time.}   Figure \ref{fig:14} shows the mass-weighted distribution of $\Delta j$ for the inflowing particles split by whether they end up as outflowing gas particles, stellar particles or remaining gas particles, for all our sample of TNG50 galaxies of the four stellar mass bins.   The histograms are normalized by the total number of particles in each category.

As can be seen, for all the four stellar mass bins, the distribution of $\Delta j$ is nearly symmetric.  The mass-weighted average is slightly negative and shows a small offset to zero, as indicated by the vertical dashed line in each panel.  These small offsets correspond to the loss of 4.8\%, 8.2\%, 8.5\% and 7.6\% of the initial $J$ for the inflowing gas particles for the four stellar mass bins, respectively.    Most strikingly, the distribution of $\Delta j$ evidently does {\it not} depend on the ultimate fate of particles, i.e. whether they later become outflowing particles or formed stars, or remain in the volume as gas particles.  In other words, there is no evidence that those gas particles that are eventually being formed into stars have preferentially lost angular momentum.

It is clear that gas particles in TNG50 simulation that flow into galaxies, or more specifically, flow through the cylindrical boundary considered here, only lose a very small fraction of angular momentum after they are accreted. In other words, there appears to be a lack of effective mechanisms to remove angular momentum of the inflowing gas in the simulations, and transport it outwards, in TNG50 simulation.  We propose that this may be the reason why the simulated star-forming disks are typically larger than the real disk galaxies for TNG50 and many hydro-dynamical simulations \citep[e.g.][]{Furlong-15, Bottrell-17, Genel-18}, as well as why the radial profiles of $\Sigma_{\rm SFR}$ for TNG50 galaxies show larger deviation from exponential function than those of observations, typically with holes in $\Sigma_{\rm SFR}$ at the center for galaxies of two highest mass bins.     

\cite{Wang-22} have proposed a disk formation model that the galactic gas disk is viewed as a ``modified accretion disk'' in which coplanar gas inflow, driven by viscous processes in the disk, provides the fuel for star formation.  In this scenario, \cite{Wang-22} found that magnetic stresses arising from magneto-rotational instability are the most plausible source of the required viscosity for the formation and maintenance of exponential star-forming disks $\Sigma_{\rm SFR}(r)$.  As with all viscous disks, the viscous stresses remove angular momentum from the inspiralling gas and transport it outwards.  Specifically, by linking the magnetic field strength to the local $\Sigma_{\rm SFR}$ ($B_{\rm tot} \propto \Sigma_{\rm SFR}^\alpha$),  \cite{Wang-22} showed that the gas disk can reach a stable steady-state with exponential $\Sigma_{\rm SFR}$ of reasonable scalelength, as long as $\alpha \sim$ 0.15, the value indicated from spatially-resolved observations of nearby galaxies \citep[e.g.][]{Tabatabaei-13, Heesen-14}.   It should be stressed that the setting up of the exponential profile in $\Sigma_{\rm SFR}$ is independent of the assumed star-formation law and thus also the gas profile. 

If this picture is correct, the size (or the angular momentum) of the resultant stellar disk is not closely connected to the angular momentum of the inflowing gas. Instead, the angular momentum of the gas disk is the result of the action of the viscous process within the disk which remove substantial amounts of angular momentum from the inflowing gas.  In this scenario, 50\%-70\% of the initial $J$ of the gas at the boundary of 10 kpc is lost by the time that gas is eventually formed into stars, for a typical disk scalelength of 2-5 kpc \citep[see figure 14 in][]{Wang-22}. This is much larger than the loss of angular momentum in the simulated galaxies in TNG50 ($<10$\%).  Transportation of angular momentum by magnetic stress is not implemented in the current magneto-hydrodynamic simulations \citep[e.g.][]{Schaye-15, Vogelsberger-14, Nelson-18}.   We suggest that doing so is potentially a way to produce smaller disks with exponential profiles of $\Sigma_{\rm SFR}$, as seen observations.

\section{Summary}
\label{sec:6}

In recent years, cosmological hydrodynamical simulations, such as EAGLE and IllustrisTNG, have made great advances in understanding galaxy formation  \citep[see][and references therein]{Vogelsberger-20}. They successfully reproduce many observational facts about the galaxy population, including galaxy colour bimodality, cold gas fraction, statistical properties of galaxy morphology and etc \citep{Trayford-15, Furlong-15, Nelson-18, Genel-18, Diemer-19, Donnari-19, Rodriguez-Gomez-19}. However, it is not clear whether these simulations are currently able to reproduce the detailed internal structure of galaxies. 

We take advantage of the publicly-released state-of-art simulation, TNG50 \citep[e.g.][]{Pillepich-18, Nelson-18}, which is the successor of the Illustris simulation with an updated physical model \citep{Vogelsberger-14, Genel-14}.  In this work, we focus on the disks and ask whether these simulated galaxies have exponential stellar and star-forming disks. IllustrisTNG has been shown to be in excellent consistent with a wide range of observational data \citep[see details in][]{Nelson-DR}.  
In addition to the general comparison between simulations and observations, in this work, we also perform a particle-level analysis of gas evolution, in order to understand whether the simulated galaxies can be treated as gas-regulator systems, and also to investigate how the angular momentum of gas particles change after they are accreted onto gas disks. 

We examine a randomly selected sample of main sequence star-forming galaxies from TNG50, excluding galaxies with significant mergers since a redshift of 0.4.   The main results of this analysis are as follows. 

\begin{itemize}

\item TNG50 star-forming galaxies tend to have larger star-forming disks, and show larger deviations from exponential profiles in the star-formation surface density $\Sigma_{\rm SFR}$ when compared with MaNGA galaxies.  Specifically,  the $\Sigma_{\rm SFR}$ profiles in TNG50 galaxies are often quite flat out to radii of 1.6$R_{1/2}$, across the range of stellar mass, with a central peak for $\log M_*/({\rm M}_{\odot}h^{-1})<10.5$, and a central suppression for galaxies with $\log M_*/({\rm M}_{\odot}h^{-1})>10.5$.  Real galaxies have a much more exponential profile in $\Sigma_{\rm SFR}$.

\item The stellar surface density profiles of TNG50 galaxies do however show good exponential profiles, in good consistency with observations.  By comparing the radial distributions of stars when they are formed at earlier epochs and of the same set of stars at the current epoch, we find that the exponential stellar disks are the result of strong radial migration of stars.  However, this strong radial migration may not be realistic for real galaxies because the radial profiles of mass-weighted age for TNG50 galaxies appear to be significantly flatter than those of real MaNGA galaxies. 

\item   By investigating the net inflow rate of individual TNG50 galaxies between the redshifts $z \sim 0.4$ and $z \sim 0$,  we find that the net inflow rate is slight less than, or comparable to, the SFR, while both are considerably larger than the absolute rate of change in the gas mass at least for galaxies with $\log M_*/({\rm M}_{\odot}h^{-1})<10.5$.   Allowing for the cosmic reduction of SFR in massive star-forming galaxies, we conclude that the star formation in TNG50 galaxies is therefore being sustained by continuous gas accretion, rather than the consumption of pre-existing gas in galaxies.  As expected, the simulated galaxies can be treated as gas-regulator systems in which gas ``flows through" the galaxies on short timescales, with attendant implications for chemical enrichment and other properties of galaxies.

\item  By tracking the evolution of individual gas particles that enter the galaxies, we find that there is no significant systematic loss of angular momentum after gas is accreted onto the disks.  Furthermore, there is no correlation between the change of angular momentum of a gas particle after it is accreted and its eventual fate, i.e. whether it is later transformed into a star-particle, is ejected from the galaxy in an outflow, or remains as a gas particle within the galaxy.  These suggest that the simulations lack an effective mechanism for removing angular momentum from the gas in the disk and transporting it outwards, and that this may account for the somewhat large sizes, and non-exponential profiles, of the star-forming disks in the simulated galaxies.   We have argued in a previous paper \citep{Wang-22} that magnetic stress from magneto-rotational instability within the disk are the most plausible source of the viscosity that is required to maintain the inwards flow of gas to produce an exponential star-forming disk.   Adding such viscosity in future simulations may potentially reduce the simulated disk sizes, and also produce more exponential star-forming disks. 

\end{itemize}

\begin{acknowledgments}
We thank the referee for their report on our paper, which has helped us improve the paper and understand better some aspects of the TNG simulations. 
E.W. acknowledges the support from the start-up Fund by University of Science and Technology of China (no. KY2030000200). 
The IllustrisTNG simulations were undertaken with compute time awarded by the Gauss Centre for Supercomputing (GCS) under GCS Large-Scale Projects GCS-ILLU and GCS-DWAR on the GCS share of the supercomputer Hazel Hen at the High Performance Computing Center Stuttgart (HLRS), as well as on the machines of the Max Planck Computing and Data Facility (MPCDF) in Garching, Germany.
\end{acknowledgments}

\bibliography{rewritebib.bib}

\end{document}